\documentclass[apj,iop]{emulateapj}

\usepackage{amsmath}
\usepackage{graphicx}
\usepackage{natbib}
\usepackage{color}

\newcommand{\snia}{SN~Ia}
\newcommand{\sneia}{SNe~Ia}

\newcommand{\ha}{H$\alpha$}

\newcommand{\CI}{C~{\sc i}}
\newcommand{\CII}{C~{\sc ii}}
\newcommand{\NaI}{Na~{\sc i}}

\newcommand{\SiII}{Si~{\sc ii}}

\newcommand{\CaII}{Ca~{\sc ii}}

\newcommand{\CrII}{Cr~{\sc ii}}

\newcommand{\FeII}{Fe~{\sc ii}}

\newcommand{\siline}{\SiII~$\lambda$6355}
\newcommand{\siblue}{\SiII~$\lambda$3858}
\newcommand{\canir}{\CaII\ IR triplet}
\newcommand{\cahk}{\CaII\ H\&K}
\newcommand{\naid}{\NaI~D}
\newcommand{\vsi}{$v_{\rm Si}$}

\newcommand{\wifes}{WiFeS}
\newcommand{\pywifes}{PyWiFeS}

\begin{document}

\title{Spectroscopic Observations of SN~2012fr: A Luminous, Normal Type Ia Supernova with Early High-Velocity Features and a Late Velocity Plateau}

\author{
    M.~J.~Childress\altaffilmark{1,2,3},
    R.~A.~Scalzo\altaffilmark{1,2}, 
    S.~A.~Sim\altaffilmark{1,2,4}, 
    B.~E.~Tucker\altaffilmark{1}, 
    F.~Yuan\altaffilmark{1,2}, 
    B.~P.~Schmidt\altaffilmark{1,2},
    S.~B.~Cenko\altaffilmark{5},
    J.~M.~Silverman\altaffilmark{6},
    C.~Contreras\altaffilmark{7}, 
    E.~Y.~Hsiao\altaffilmark{7}, 
    M.~Phillips\altaffilmark{7}, 
    N.~Morrell\altaffilmark{7}, 
    S.~W.~Jha\altaffilmark{8},
    C.~McCully\altaffilmark{8},
    A.~V.~Filippenko\altaffilmark{5},
    J.~P.~Anderson\altaffilmark{9},
    S.~Benetti\altaffilmark{10},
    F.~Bufano\altaffilmark{11},
    T.~de~Jaeger\altaffilmark{9},
    F.~Forster\altaffilmark{9},
    A.~Gal-Yam\altaffilmark{12},
    L.~Le~Guillou\altaffilmark{13},
    K.~Maguire\altaffilmark{14},
    J.~Maund\altaffilmark{4},
    P.~A.~Mazzali\altaffilmark{15,10,16},
    G.~Pignata\altaffilmark{11},
    S.~Smartt\altaffilmark{4},
    J.~Spyromilio\altaffilmark{17},
    M.~Sullivan\altaffilmark{18},
    F.~Taddia\altaffilmark{19},
    S.~Valenti\altaffilmark{20,21},
    D.~D.~R.~Bayliss\altaffilmark{1}, 
    M.~Bessell\altaffilmark{1},
    G.~A.~Blanc\altaffilmark{22},
    D.~J.~Carson\altaffilmark{23},
    K.~I.~Clubb\altaffilmark{5},
    C.~de~Burgh-Day\altaffilmark{24},
    T.~D.~Desjardins\altaffilmark{25},
    J.~J.~Fang\altaffilmark{26},
    O.~D.~Fox\altaffilmark{5},
    E.~L.~Gates\altaffilmark{26},
    I-T.~Ho\altaffilmark{27},
    S.~Keller\altaffilmark{1}, 
    P.~L.~Kelly\altaffilmark{5},
    C.~Lidman\altaffilmark{28},
    N.~S.~Loaring\altaffilmark{29},
    J.~R.~Mould\altaffilmark{30},
    M.~Owers\altaffilmark{28},
    S.~Ozbilgen\altaffilmark{24},
    L.~Pei\altaffilmark{23},
    T.~Pickering\altaffilmark{29},
    M.~B.~Pracy\altaffilmark{31},
    J.~A.~Rich\altaffilmark{22},
    B.~E.~Schaefer\altaffilmark{32},
    N.~Scott\altaffilmark{30},
    M.~Stritzinger\altaffilmark{33},
    F.~P.~A.~Vogt\altaffilmark{1},
    G.~Zhou\altaffilmark{1}
}

\altaffiltext{1}
{
    Research School of Astronomy and Astrophysics, 
    Australian National University, 
    Canberra, ACT 2611, Australia.
}
\altaffiltext{2}
{
    ARC Centre of Excellence for All-sky Astrophysics (CAASTRO).
}
\altaffiltext{3}
{
    E-mail: mjc@mso.anu.edu.au .
}
\altaffiltext{4}
{
    Astrophysics Research Centre, 
    School of Mathematics and Physics, 
    Queen's University Belfast, 
    Belfast BT7 1NN, UK.
}
\altaffiltext{5}
{
    Department of Astronomy, 
    University of California, 
    Berkeley, CA 94720-3411, USA.
}
\altaffiltext{6}
{
    Department of Astronomy, 
    University of Texas, 
    Austin, TX 78712-0259, USA.
}
\altaffiltext{7}
{
    Las Campanas Observatory, 
    Carnegie Observatories, Casilla 601, 
    La Serena, Chile.
}
\altaffiltext{8}
{
    Department of Physics and Astronomy, 
    Rutgers, the State University of New Jersey, 
    136 Frelinghuysen Road, 
    Piscataway, NJ 08854, USA.
}
\altaffiltext{9}
{
    Departamento de Astronom\'ia, 
    Universidad de Chile, 
    Casilla 36-D, Santiago, Chile.
}
\altaffiltext{10}
{
    INAF Osservatorio Astronomico di Padova, 
    Vicolo dell'Osservatorio 5, 
    35122 Padova, Italy.
}
\altaffiltext{11}
{
    Departamento de Ciencias Fisicas, 
    Universidad Andres Bello, Avda. Republica 252, 
    Santiago, Chile.
}
\altaffiltext{12}
{
    Department of Particle Physics and Astrophysics, 
    The Weizmann Institute of Science, 
    Rehovot 76100, Israel.
}
\altaffiltext{13}
{
    UPMC Univ. Paris 06, UMR 7585, 
    Laboratoire de Physique Nucleaire et des
    Hautes Energies (LPNHE), 75005 Paris, France.
}
\altaffiltext{14}
{
    Department of Physics (Astrophysics), 
    University of Oxford, DWB, Keble Road, 
    Oxford OX1 3RH, UK.
}
\altaffiltext{15}
{
    Astrophysics Research Institute, 
    Liverpool John Moores University,
    Egerton Wharf, Birkenhead, CH41 1LD, UK.
}
\altaffiltext{16}
{
    Max-Planck-Institut f\"ur Astrophysik, 
    Karl-Schwarzschild str. 1, 85748 Garching, Germany.
}
\altaffiltext{17}
{
    European Southern Observatory,
    Karl-Schwarzschild-Strasse 2,
    Garching D-85748, Germany.
}
\altaffiltext{18}
{
    School of Physics and Astronomy, 
    University of Southampton, 
    Southampton SO17 1BJ, UK.
}
\altaffiltext{19}
{
    The Oskar Klein Centre, 
    Department of Astronomy, AlbaNova, 
    Stockholm University, 10691 Stockholm, Sweden.
}
\altaffiltext{20}
{
    Las Cumbres Observatory Global Telescope Network, 
    6740 Cortona Dr., Suite 102, 
    Goleta, CA 93117, USA.
}
\altaffiltext{21}
{
    Department of Physics, 
    University of California, 
    Broida Hall, Mail Code 9530, 
    Santa Barbara, CA 93106-9530, USA.
}
\altaffiltext{22}
{
    Observatories of the Carnegie Institution of Washington,
    813 Santa Barbara St.,
    Pasadena, CA 91101, USA.
}
\
\altaffiltext{23}
{
    Department of Physics and Astronomy, 
    University of California, 
    Irvine, CA 92697-4575, USA.
}
\altaffiltext{24}
{
    School of Physics, 
    University of Melbourne, 
    Parkville, VIC 3010, Australia.
}
\altaffiltext{25}
{
    Department of Physics and Astronomy, 
    The University of Western Ontario, 
    London, ON N6A 3K7, Canada.
}
\altaffiltext{26}
{
    University of California Observatories/Lick Observatory,
    University of California, 
    Santa Cruz, CA 95064, USA.
}
\altaffiltext{27}
{
    Institute for Astronomy, 
    University of Hawaii, 
    2680 Woodlawn Drive, 
    Honolulu, HI 96822, USA.
}
\altaffiltext{28}
{
    Australian Astronomical Observatory, 
    PO Box 915, North Ryde, NSW 1670, Australia.
}
\altaffiltext{29}
{
    South African Astronomical Observatory (SAAO),
    P.O. Box 9, Observatory 7935, South Africa.
}
\altaffiltext{30}
{
    Centre for Astrophysics \& Supercomputing, 
    Swinburne University of Technology, 
    PO Box 218, Hawthorn, VIC 3122, Australia.
}
\altaffiltext{31}
{
    Sydney Institute for Astronomy, 
    School of Physics, 
    University of Sydney, NSW 2006, Australia.
}
\altaffiltext{32}
{
    Department of Physics and Astronomy, 
    Louisiana State University, 
    Baton Rouge, LA 70803, USA.
}
\altaffiltext{33}
{
    Department of Physics and Astronomy, 
    Aarhus University, Ny Munkegade 120, DK-8000 Aarhus C, Denmark.
}

\begin{abstract}
We present 65 optical spectra of the Type Ia SN 2012fr,
of which 33 were obtained before maximum light.  
At early times SN~2012fr shows clear evidence of a high-velocity feature (HVF) in the \siline\ line which can be cleanly decoupled from the lower velocity ``photospheric'' component. This \siline\ HVF fades by phase $-5$; subsequently, the photospheric component exhibits a very narrow velocity width and remains at a nearly constant velocity of $\sim$12,000~km~s$^{-1}$ until at least 5 weeks after maximum brightness. The \CaII\ infrared (IR) triplet exhibits similar evidence for both a photospheric component at $v \approx$ 12,000~km~s$^{-1}$ with narrow line width and long velocity plateau, as well as a high-velocity component beginning at $v \approx$ 31,000~km~s$^{-1}$ two weeks before maximum.
SN~2012fr resides on the border between the ``shallow silicon'' and ``core-normal'' subclasses in the \citet{branch09} classification scheme, and on the border between normal and ``high-velocity'' \sneia\ in the \citet{wang09} system. Though it is a clear member of the ``low velocity gradient'' \citep[LVG;][]{benetti05} group of \sneia\ and exhibits a very slow light-curve decline, it shows key dissimilarities with the overluminous SN~1991T or SN~1999aa subclasses of \sneia. 
SN~2012fr represents a well-observed \snia\ at the luminous end of the normal \snia\ distribution, and a key transitional event between nominal spectroscopic subclasses of \sneia.
\end{abstract}


\section{Introduction}
Type Ia supernovae (\sneia) are critical cosmological tools for measuring the expansion history of the Universe \citep{riess98, perlmutter99}, yet much remains unknown about the nature of these enlightening explosions. Their luminosities show low intrinsic dispersion ($\sim0.35$~mag) and they generally obey a scaling of their absolute luminosity with the width of their optical light curve \citep{phillips93, phillips99}, about which the brightness dispersion is even lower. The width-luminosity relationship appears to be driven by the amount of radioactive $^{56}$Ni produced in the explosion and the opacity \citep{hk96, pe00, mazzali01, mazzali07}, but the progenitor mechanism driving these properties remains uncertain. 
While it is generally accepted that \sneia\ arise from the thermonuclear disruption of a carbon-oxygen (C-O) white dwarf (WD) in a binary system \citep{hf60}, scenarios in which the companion is a main sequence or red giant star \citep[the single-degenerate scenario;][]{whelan73} and those in which the companion is another WD \citep[the double-degenerate scenario;][]{tutukov76, tutukov79, iben84, webbink84} have both proven consistent with some observational features of \sneia.
Whether \sneia\ represent a unified class of objects with a common physical origin or result from multiple progenitor channels has yet to be determined, and is a critical question for the continued use of \sneia\ in cosmology.

Optical spectroscopy of \sneia\ can provide vital insight into the question of \snia\ diversity. Large samples of \snia\ spectra have been made publicly available \citep[e.g.,][]{matheson08, blondin12, bsnip1, wiserep}, and investigations of spectroscopic subclassification of \sneia\ have been a vigorous area of study \citep{benetti05, branch09, wang09, bsnip2}. While a rigorous accounting of the diversity of \sneia\ is crucial for understanding the source of their luminosity dispersion, individual cases of well-studied \sneia\ \citep[e.g.,][]{stanishev03du, wang05cf, foley09ig, silverman12cg} can yield key insights into the nature of the explosions themselves.

In this work we focus on optical spectroscopy of SN~2012fr, a \snia\ which was discovered on 2012 Oct. 27 in the nearby barred spiral galaxy NGC~1365. Shortly after its discovery we initiated a rigorous photometric and spectroscopic follow-up program for SN~2012fr. Optical photometry will be presented by Contreras et al. (2013, hereafter Paper~II), where we show that SN~2012fr has a normal light curve for an \snia. This paper presents optical spectra of SN~2012fr, while Hsiao et al. (2013, Paper III) will present near-infrared (NIR) spectra and Tucker et al. (2013, Paper IV) will analyze constraints on the progenitor from pre-explosion imaging and very early photometry.

This paper is organized as follows. In \S~\ref{sec:data} we present the observational data. Section~\ref{sec:si_line} focuses on the \siline\ line, and characterizes both the high-velocity features observed at early times and the long velocity plateau observed at late times. Other absorption features of particular note for SN~2012fr -- including narrow \naid, unburned C, the \canir, and Fe-group elements -- are inspected in \S~\ref{sec:other_lines}. We address the spectroscopic ``subclassification'' of SN~2012fr in the context of modern classification schemes in \S~\ref{sec:classification}. We then discuss implications of our observational results in \S~\ref{sec:discussion} and present concluding remarks in \S~\ref{sec:conclusions}.

\section{Spectroscopic Observations}
\label{sec:data}
SN~2012fr was discovered on 2012 Oct. 27 (UT dates are used throughout this paper) by \citet{klotz12fr} at $\alpha=$ 03$^{\rm h}$33$^{\rm m}$36.274$^{\rm s}$, $\delta = -36^\circ07'34.46''$ (J2000) in the nearby barred spiral galaxy NGC~1365, and shortly thereafter classified as a SN~Ia \citep{childress12frcbet, buil12fr}. Extensive photometric coverage presented in Paper~II shows that SN~2012fr reached a peak brightness of $m_B=12.0$ mag on 2012 Nov. 12.04 with a 15-day decline of $\Delta m_{15}(B)=0.80$ mag. Given the nominal distance modulus to NGC~1365 of $\mu=31.3$ mag \citep{silbermann99, freedman01}, this implies a peak luminosity of $M_B=-19.3$ mag, placing it in firm agreement with the \citet{phillips93} relation.

Spectra of SN~2012fr were collected at multiple locations.  The two main sources were the Wide Field Spectrograph \citep[\wifes;][]{dopita07, dopita10} on the Australian National University (ANU) 2.3 m telescope at Siding Spring Observatory in northern New South Wales, Australia, and the Public ESO Spectroscopic Survey of Transient Objects (PESSTO) utilizing the 3.6 m New Technology Telescope (NTT) at La Silla, Chile.

\wifes\ spectra were obtained using the B3000 and R3000 gratings, providing wavelength coverage from 3500 \AA\ to 9600 \AA\ with a resolution of 1.5 \AA\ and 2.5 \AA\ (all reported instrument resolutions are full width at half-maximum intensity, FWHM) in the blue and red channels, respectively. Data cubes for \wifes\ observations were produced using the \pywifes\ software\footnote[34]{{\tt http://www.mso.anu.edu.au/pywifes/ .}}
(Childress et al. 2013, in prep.).  
Spectra of the SN were extracted from final data cubes using a point-spread function (PSF) weighted extraction technique with a simple symmetric Gaussian PSF, and the width of this Gaussian was measured directly from the data cube.  We found this method to produce flux measurements consistent with a simple aperture extraction method, but with improved signal-to-noise ratio. Background subtraction was performed by calculating the median background spectrum across all spaxels outside a distance from the SN equal to about three times the seeing disk (which was typically 1.5--2\arcsec\ FWHM).  Due to the negligible galaxy background and good spatial flatfielding from the \pywifes\ pipeline, this technique produced favorable subtraction of the sky background from the \wifes\ spectra of SN~2012fr.

A major component of our observing campaign was a series of optical and NIR spectra obtained as part of the PESSTO (Smartt et al. 2013, in prep.)\footnote[35]{{\tt http://www.pessto.org .}} survey using the NTT-3.6 m telescope in La Silla, Chile.  
Optical spectra from PESSTO were obtained with EFOSC2 \citep{efosc} using the Gr11 and Gr16 grisms, which both have a resolution of 16~\AA.  
NIR spectra were obtained with SOFI \citep{sofi} using the GB and GR grisms, which give respective resolutions of 14\AA\ and 21\AA, with observations dithered to facilitate sky background subtraction. SOFI spectra will be released as part of the PESSTO data products for SN~2012fr, and will constitute a portion of the NIR spectra of SN~2012fr analyzed in Paper~III.
Both EFOSC and SOFI spectra were reduced using the PESSTO pipeline developed by S. Valenti, which is a custom-built python/pyraf package that performs all standard spectroscopic reduction steps including preprocessing, wavelength and flux calibration, spectrum extraction, and removal of telluric features measured from long exposure standard star spectra.
Continued observations of SN~2012fr in 2013 are ongoing as part of the PESSTO operations and will be presented in a future PESSTO paper.


Additional spectra of SN~2012fr were obtained with the Robert Stobie Spectrograph on the South African Large Telescope (SALT), the Grating Spectrograph on the South African Astronomical Observatory (SAAO) 1.9 m telescope, the Kast Double Spectrograph \citep{kast} on the Shane 3 m telescope at Lick Observatory, the Wide Field Reimaging CCD Camera (WFCCD) on the 2.5 m Ir\'en\'ee du Pont telescope at Las Campanas Observatory, the Inamori-Magellan Areal Camera and Spectrograph \citep[IMACS;][]{imacs} on the 6 m Magellan-Baade telescope at Las Campanas, and the Andalucia Faint Object Spectrograph and Camera (ALFOSC) on the 2.5 m Nordic Optical Telescope (NOT) on La Palma.

SALT/RSS observations were obtained with a 900 l\,mm$^{-1}$ VPH grating at three tilt angles to cover the range 3480--9030~\AA. The 1.5\arcsec\-wide slit yielded a resolution of $\sim 6$~\AA. Initial processing of the SALT data utilized the SALT science pipeline PySALT \footnote[36]{{\tt http://pysalt.salt.ac.za .}} \citep{crawford10}.
SAAO-1.9 m observations used the 300 l\,mm$^{-1}$ grating (\#7) at an angle of 17.5$^\circ$, corresponding to a central wavelength of 5400~\AA, a wavelength range of $\sim 3500$--7300~\AA, and a resolution of 5~\AA.
%
Lick/Kast observations employed the 600 l\,mm$^{-1}$ grating on the
blue arm, blazed at 4310~\AA, and provides wavelength coverage of
3500--5600~\AA\ with a resolution of 6--7~\AA. Different observers
used different gratings on the Kast red arm, including the 300
l\,mm$^{-1}$ grating blazed at 7500~\AA\ and covering
5500--10,300~\AA\ with a resolution of 11~\AA, the 600 l\,mm$^{-1}$
grating blazed at 7000~\AA\ and covering 5600--8200~\AA\ with
5.5~\AA\ resolution, and the 830 l\,mm$^{-1}$ grating blazed at
6500~\AA\ and covering 5600--7440~\AA\ with 4~\AA\ resolution.
%
WFCCD observations were obtained with the 400 l\,mm$^{-1}$ grism yielding 8~\AA\ resolution, and data were reduced following the procedures described in detail by \citet{hamuy06}. The IMACS spectrum employed the 300 l\,mm$^{-1}$ grating and 0.9\arcsec\ slit yielding a resolution of 2.7~\AA. 

All long-slit low-resolution spectra were reduced using standard
techniques \citep[e.g.,][]{foley03}. Routine CCD processing and
spectrum extraction were completed with IRAF. We obtained the
wavelength scale from low-order polynomial fits to calibration-lamp
spectra. Also, we fit a spectrophotometric standard-star spectrum to
the data in order to flux calibrate the SN and to remove telluric
absorption lines.

We obtained a high-resolution optical spectrum of SN~2012fr with the High Resolution Echelle Spectrometer \citep[HIRES;][]{hires} on the 10 m Keck I telescope with the blue cross-disperser (``HIRESb'') on 2012 Oct. 29.45.  We used the C2 decker (i.e., the 1.15\arcsec\ slit), providing coverage from the atmospheric cutoff to $\lambda = 5960$~\AA\ with a resolution of 37,000.

\begin{figure}
\begin{center}
\includegraphics[width=0.45\textwidth]{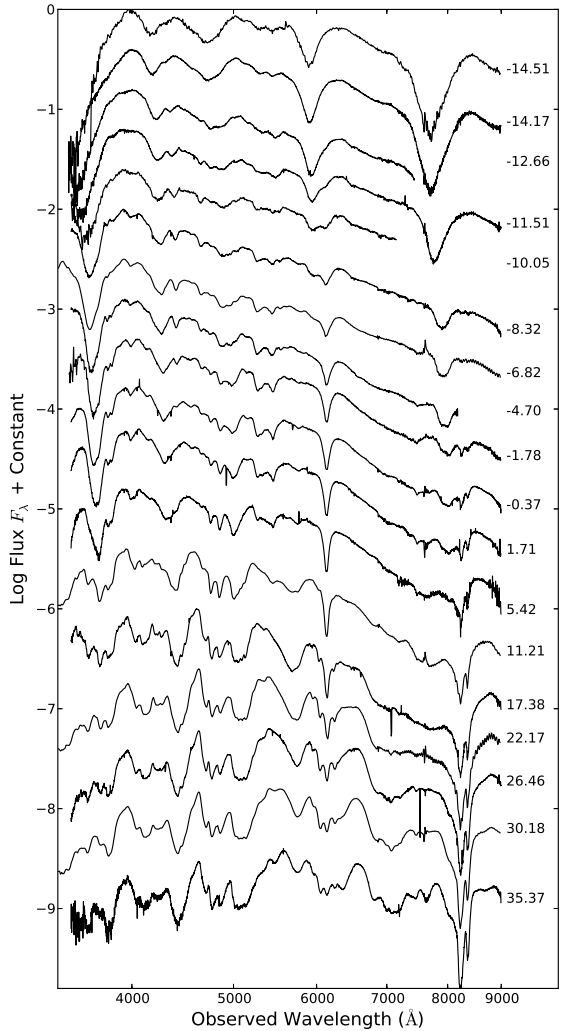}
\caption{Representative sample of SN~2012fr spectra, labeled by phase with respective to $B$-band maximum light.}
\label{fig:time_series}
\end{center}
\end{figure}

A full table of our optical spectra is given in Table~\ref{tab:obs_log}, and a representative plot of our spectral time series is shown in Figure~\ref{fig:time_series}.
At the earliest epochs of SN~2012fr, our observing strategy was to request spectra from multiple sources worldwide. This resulted in several spectra during the same night (often separated by 0.3--0.5 day) on some occasions, but consistently resulted in at least one spectrum every night until nearly two weeks after maximum light.
On nights with extremely poor seeing ($>$3\arcsec) at Siding Spring, some \wifes\ observers chose to observe SN~2012fr multiple times due to the inability to observe their own fainter targets.
Upon publication of this paper we will make all of our optical spectra publicly available via the WISEREP \citep{wiserep} SN spectroscopy repository.

\begin{table*}
\scriptsize
\begin{center}
\caption{Optical Spectroscopy Observation Log}
\label{tab:obs_log}
\begin{tabular}{lccrcl}
\hline
UT Date      & Phase$^a$ & Telescope        & Exposure & Wavelength     & Observers$^b$ \\
             & (days)    & / Instrument     & Time (s) & Range (\AA)   &               \\
\hline
2012-Oct-28.53 & -14.51  & ANU-2.3m / WiFeS &  900     & 3500-9550     & GZ, DB \\
2012-Oct-28.87 & -14.17  & SALT / RSS       & 1200     & 3480-9030     & SJ, CM1 \\
2012-Oct-29.45 & -13.59  & Keck-I / HIRES   & 1200     & 3500-5960     & BZ, MJ, SX, BK \\
2012-Oct-30.38 & -12.66  & Lick-3m / Kast   & 2000     & 3500-7440     & CM2, BZ \\
2012-Oct-30.51 & -12.53  & ANU-2.3m / WiFeS & 1200     & 3500-9550     & GZ, DB \\
2012-Oct-31.53 & -11.51  & ANU-2.3m / WiFeS & 1200     & 3500-9550     & GZ, DB \\
2012-Nov-01.59 & -10.45  & ANU-2.3m / WiFeS & 1200     & 3500-5700$^c$ & MB, SK \\
2012-Nov-01.99 & -10.05  & SAAO-1.9m / GS   &  900     & 3500-7150     & NSL \\
2012-Nov-02.48 &  -9.56  & ANU-2.3m / WiFeS &  900     & 3500-9550     & MB, SK \\
2012-Nov-02.69 &  -9.35  & ANU-2.3m / WiFeS &  900     & 3500-9550     & MB, SK \\
2012-Nov-03.05 &  -8.99  & SAAO-1.9m / GS   &  900     & 3500-7150     & NSL \\
2012-Nov-03.57 &  -8.47  & ANU-2.3m / WiFeS &  900     & 3500-9550     & MB, SK \\
2012-Nov-03.72 &  -8.32  & ANU-2.3m / WiFeS &  900     & 3500-9550     & MB, SK \\
2012-Nov-04.07 &  -7.97  & SAAO-1.9m / GS   &  900     & 3500-7150     & NSL \\
2012-Nov-04.34 &  -7.70  & Lick-3m / Kast   &  600     & 3500-8220     & EG \\
2012-Nov-04.50 &  -7.54  & ANU-2.3m / WiFeS &  900     & 3500-9550     & MB, SK \\
2012-Nov-05.22 &  -6.82  & NTT-3.6m / EFOSC &  100     & 3360-10000    & PESSTO \\
2012-Nov-05.37 &  -6.67  & Lick-3m / Kast   &  600     & 3500-8220     & EG \\
2012-Nov-05.61 &  -6.43  & ANU-2.3m / WiFeS &  900     & 3500-9550     & MB, SK \\
2012-Nov-06.38 &  -5.66  & Lick-3m / Kast   &  180     & 3500-10300    & SBC, PK \\
2012-Nov-07.27 &  -4.77  & NTT-3.6m / EFOSC &  100     & 3360-10000    & PESSTO \\
2012-Nov-07.28 &  -4.76  & du Pont / WFCCD  &   60     & 3500-9600     & NM, BM \\
2012-Nov-07.34 &  -4.70  & Lick-3m / Kast   &  180     & 3500-8180     & LP, DC \\
2012-Nov-08.24 &  -3.80  & NTT-3.6m / EFOSC &  100     & 3360-10000    & PESSTO \\
2012-Nov-08.27 &  -3.77  & du Pont / WFCCD  &   60     & 3500-9600     & NM, BM \\
2012-Nov-08.34 &  -3.70  & Lick-3m / Kast   &  180     & 3500-8180     & LP, DC \\
2012-Nov-09.52 &  -2.52  & ANU-2.3m / WiFeS &  900     & 3500-9550     & FV \\
2012-Nov-09.62 &  -2.42  & ANU-2.3m / WiFeS &  900     & 3500-9550     & FV \\
2012-Nov-10.26 &  -1.78  & du Pont / WFCCD  &   80     & 3500-9600     & NM, BM \\
2012-Nov-10.57 &  -1.47  & ANU-2.3m / WiFeS &  700     & 3500-9550     & FV \\
2012-Nov-10.70 &  -1.34  & ANU-2.3m / WiFeS &  900     & 3500-9550     & FV \\
2012-Nov-11.26 &  -0.78  & du Pont / WFCCD  &   90     & 3500-9600     & NM, BM \\
2012-Nov-11.67 &  -0.37  & ANU-2.3m / WiFeS &  900     & 3500-9550     & FV \\
2012-Nov-12.38 &  +0.34  & Lick-3m / Kast   &  180     & 3500-8220     & TD, JF \\
2012-Nov-12.74 &  +0.70  & ANU-2.3m / WiFeS &  600     & 3500-9550     & MO, MP \\
2012-Nov-13.24 &  +1.20  & du Pont / WFCCD  &  270     & 3500-9600     & JR, GB \\
2012-Nov-13.26 &  +1.22  & NTT-3.6m / EFOSC &  100     & 3360-10000    & PESSTO \\
2012-Nov-13.75 &  +1.71  & ANU-2.3m / WiFeS &  600     & 3500-9550     & MO, MP \\
2012-Nov-14.25 &  +2.21  & du Pont / WFCCD  &  270     & 3500-9600     & JR, GB \\
2012-Nov-14.32 &  +2.28  & Lick-3m / Kast   &  300     & 3500-10300    & KC, OF \\
2012-Nov-15.22 &  +3.18  & NTT-3.6m / EFOSC &  100     & 3360-10000    & PESSTO \\
2012-Nov-15.24 &  +3.20  & du Pont / WFCCD  &  270     & 3500-9600     & JR, GB \\
2012-Nov-16.24 &  +4.20  & du Pont / WFCCD  &  270     & 3500-9600     & JR, GB \\
2012-Nov-16.51 &  +4.47  & ANU-2.3m / WiFeS &  600     & 3500-9550     & NS \\
2012-Nov-17.22 &  +5.18  & du Pont / WFCCD  &  270     & 3500-9600     & JR, GB \\
2012-Nov-17.46 &  +5.42  & ANU-2.3m / WiFeS &  600     & 3500-9550     & NS \\
2012-Nov-18.24 &  +6.20  & du Pont / WFCCD  &  270     & 3500-9600     & JR, GB \\
2012-Nov-18.42 &  +6.38  & ANU-2.3m / WiFeS &  600     & 3500-9550     & NS \\
2012-Nov-19.16 &  +7.12  & du Pont / WFCCD  &  100     & 3500-9600     & JR, GB \\
2012-Nov-19.56 &  +7.52  & ANU-2.3m / WiFeS &  900     & 3500-9550     & MC \\
2012-Nov-20.15 &  +8.11  & du Pont / WFCCD  &  100     & 3500-9600     & NM \\
2012-Nov-20.30 &  +8.26  & Lick-3m / Kast   &  360     & 3500-10300    & SBC, OF \\
2012-Nov-20.48 &  +8.44  & ANU-2.3m / WiFeS &  900     & 3500-9550     & MC \\
2012-Nov-21.13 &  +9.09  & du Pont / WFCCD  &  300     & 3500-9600     & NM \\
2012-Nov-21.24 &  +9.20  & NTT-3.6m / EFOSC &  100     & 3360-10000    & PESSTO \\
2012-Nov-21.68 &  +9.64  & ANU-2.3m / WiFeS &  900     & 3500-9550     & MC \\
2012-Nov-23.25 & +11.21  & NTT-3.6m / EFOSC &  100     & 3360-10000    & PESSTO \\
2012-Nov-29.42 & +17.38  & ANU-2.3m / WiFeS &  900     & 3500-9550     & CL, BS \\
2012-Nov-30.08 & +18.04  & Baade / IMACS    &  900     & 3400-9600     & DO \\
2012-Dec-04.21 & +22.17  & NTT-3.6m / EFOSC &  300     & 3360-10000    & PESSTO \\
2012-Dec-08.50 & +26.46  & ANU-2.3m / WiFeS &  900     & 3500-9550     & ITH \\
2012-Dec-12.22 & +30.18  & NTT-3.6m / EFOSC &  600     & 3360-10000    & PESSTO \\
2012-Dec-16.92 & +34.88  & NOT / ALFOSC     &  300     & 3300-9100     & MS \\
2012-Dec-17.41 & +35.37  & ANU-2.3m / WiFeS &  900     & 3500-9550     & JM, CD, SO \\
2012-Dec-21.22 & +39.18  & NTT-3.6m / EFOSC &  900     & 3360-10000    & PESSTO \\
\hline
\end{tabular}
\end{center}
$^a$ With respect to $B$-band maximum brightness on 2012 Nov. 12.04. \\
$^b$ 
BK, Beth Klein;
BM, Barry Madore;
BS, Brad Schaefer;
BZ, Ben Zuckerman;
CD, Catherine de Burgh-Day;
CL, Chris Lidman;
CM1, Curtis McCully;
CM2, Carl Melis;
DB, Daniel Bayliss; 
DC, Dan Carson;
DO, David Osip;
EG, Elinor Gates;
FV, Fr\'ed\'eric Vogt;
GB, Guillermo Blanc;
GZ, George Zhou;
ITH, I-Ting Ho;
JF, Jerome Fang;
JM, Jeremy Mould,
JR, Jeff Rich;
KC, Kelsey Clubb;
LP, Liuyi Pei;
MB, Mike Bessell;
MC, Mike Childress;
MJ, Michael Jura;
MO, Matt Owers;
MP, Mike Pracy;
MS, Max Stritzinger;
NM, Nidia Morrell;
NS, Nic Scott;
NSL, Nicola S. Loaring;
OF, Ori Fox;
PK, Pat Kellyl
SBC, S. Bradley Cenko;
SJ, Saurabh Jha;
SK, Stefan Keller;
SO, Sinem Ozbilgen;
SX, Siyi Xu;
TD, Tyler Desjardins.
\\
$^c$ WiFeS red channel cryo pump failure.
\normalsize
\end{table*}

%

\section{Evolution of the \SiII\ $\lambda$6355 Line}
\label{sec:si_line}
The \siline\ line is among the most prominent features in \snia\ spectra.  Both its characteristics at maximum light \citep[e.g.,][]{nugent95} and its evolution in time \citep[e.g.,][]{benetti05} have been key tools in characterizing \snia\ diversity. Additionally, the properties of this line in combination with behavior of other lines (``spectral indicators'') have been used to identify potential subclasses of \sneia\ \citep[e.g.,][]{benetti05, bongard06, branch09, bsnip2, blondin12}.

For SN~2012fr, the evolution of the \siline\ line has three key features of note: (1) the velocity width of the line (and indeed other lines; see \S~\ref{sec:other_lines}) is extremely narrow, starting about a week after maximum brightness; (2) early-time spectra show clear signatures of a two-component \siline, indicating a layer of ejected material at higher velocities than the nominal photospheric layer; and (3) the velocity of the photospheric component remains constant (to within $\sim 200$ km~s$^{-1}$) until at least $\sim 40$ days after maximum.

\begin{figure}
\begin{center}
\includegraphics[width=0.45\textwidth]{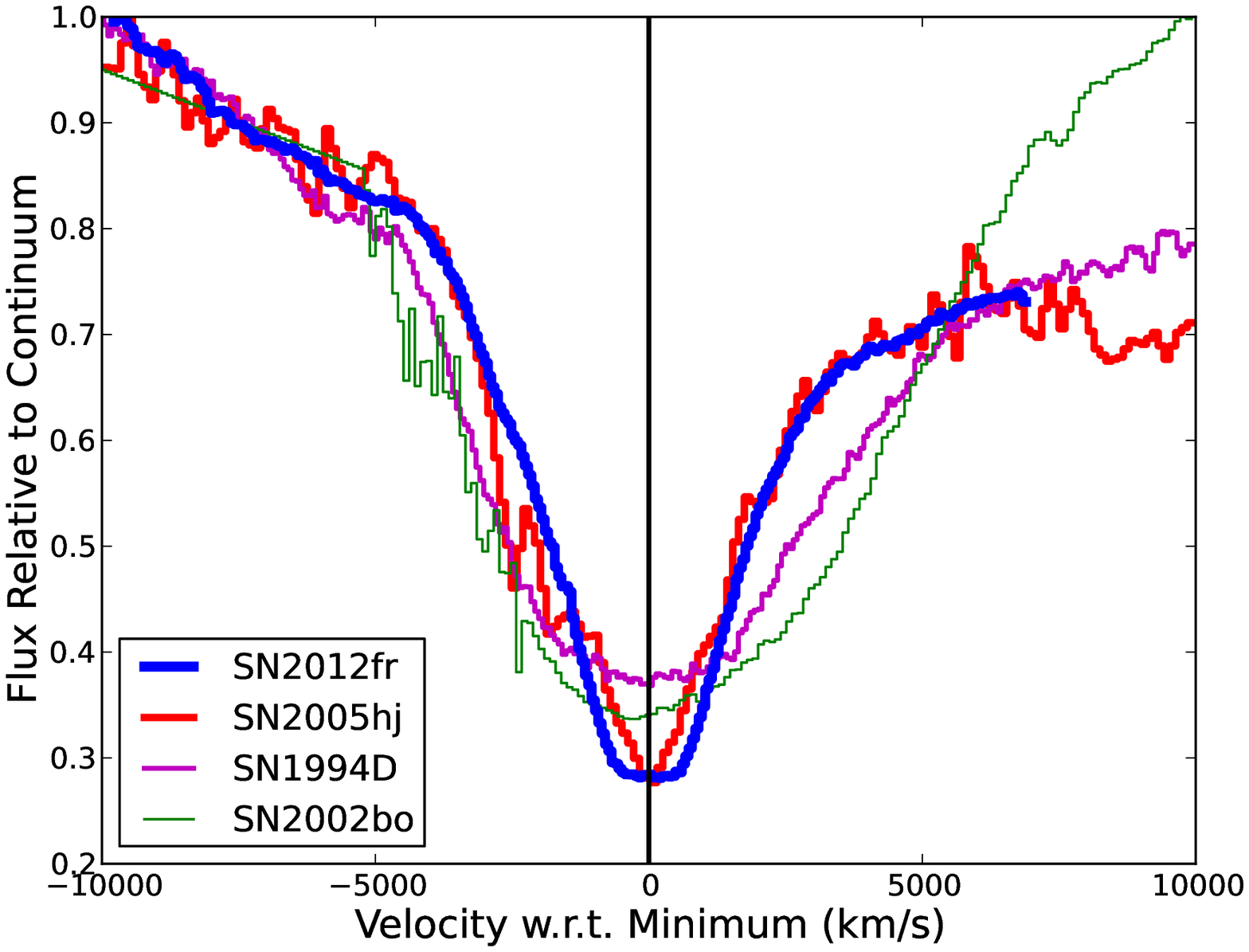}
\caption{The \siline\ feature of SN~2012fr (thick blue line) at +8 days in velocity space centered at the velocity minimum. Plotted for comparison are SN~2005hj \citep{quimby07} at +9 days in red, SN~1994D at +11 days \citep[from][]{blondin12} in magenta, and SN~2002bo \citep{benetti04} at +5 days in green, in order of thickest to thinnest lines.}
\label{fig:narrow_line}
\end{center}
\end{figure}

The clear detections of both the high-velocity layer and the constant photospheric velocity are facilitated by the extremely narrow velocity width of the photospheric absorption lines in SN~2012fr. In Figure~\ref{fig:narrow_line} we show the \siline\ width of SN~2012fr viewed in velocity space, compared to that of SN~2005hj \citep{quimby07}, SN~1994D \citep[from][]{blondin12}, and SN~2002bo \citep{benetti04}. SN~2012fr has narrower \siline\ than the other \sneia, except for perhaps SN~2005hj whose similarly narrow line width was highlighted by \citet{quimby07}.  One can even visibly identify a flattening at the base of this feature due to the doublet nature of the line. We measure the observed line width (FWHM) to be $\sim 3400$ km~s$^{-1}$; if we account for the 14~\AA\ separation of the doublet lines, this implies an intrinsic line width of $\sim 3000$ km~s$^{-1}$.

\subsection{High-Velocity \siline\ in Early-Time Spectra}
\label{sec:hvf_si}
At two weeks before maximum light, \siline\ appears to be composed of a single broad, high-velocity component, but beginning around $-12$ days a second distinct component at lower velocities begins to develop. By $-9$ days the HVF and the lower velocity component exhibit equal strength, but by $-5$ days the HVF becomes difficult to distinguish visually.

While such \siline\ HVFs have been observed in other \sneia, notably SN~2005cf \citep{wang05cf} and SN~2009ig \citep{foley09ig, marion13} (see also \S~\ref{sec:silicon_velocities}), the distinction between HVF and photospheric components is cleaner in SN~2012fr than ever seen before. In this section we follow the evolution of the two components in a quantitative way by fitting the \siline\ line as a simple double-Gaussian profile. 

\begin{figure}
\begin{center}
\includegraphics[width=0.45\textwidth]{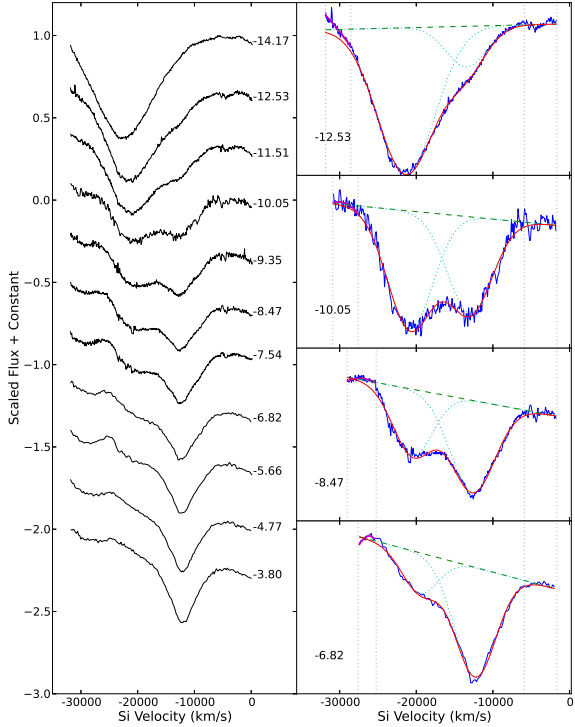}
\caption{Left: Evolution of the \siline\ line of SN~2012fr in velocity space. Right: Several representative examples of the two-component Gaussian fits. Data are in blue, regions of the pseudo-continuum fit are denoted by the vertical dotted black lines, the fitted pseudo-continuum is the dashed green line, the fully fitted profile is shown as the smooth red line, and the two individual components are the dotted cyan curves.}
\label{fig:twopeak_si_fits}
\end{center}
\end{figure}

We show in Figure~\ref{fig:twopeak_si_fits} some example fits of the \siline\ line at several epochs. We first begin by defining regions of the blue and red pseudo-continuum, then perform a simple linear fit between the two regions. The flux in the line region is next divided by the pseudo-continuum, and the normalized absorption profile is fitted with two Gaussians. The fit parameters are the center, width, and depth of each component, and the only constraints imposed are that the HVF component be above 14,000 km~s$^{-1}$ and the low-velocity photospheric component be below that same threshold. This threshold between the two fitted velocity components was chosen because it is higher than the velocities observed in most \sneia, and provided favorable separation between the two velocity components.

\begin{figure}
\begin{center}
\includegraphics[width=0.45\textwidth]{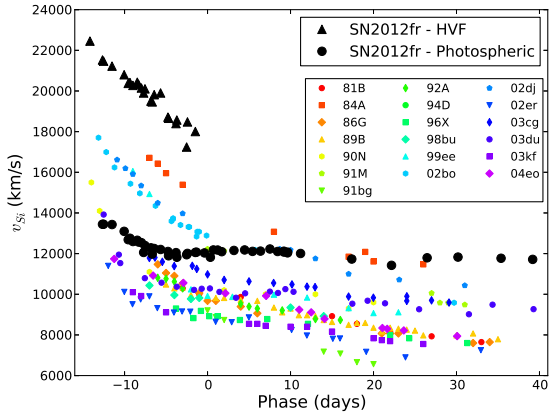}
\caption{Velocity evolution of the \siline\ feature of SN~2012fr compared to that of a number of \sneia\ from the \citet{benetti05} sample. The HVF component of SN~2012fr is shown as the large black triangles, while the lower velocity photospheric component is shown as large black circles.}
\label{fig:all_vsi}
\end{center}
\end{figure}

In Table~\ref{tab:vsi_fits} we present the fitted parameters for our two-component \siline\ fits including velocity center ($v$), velocity width ($\Delta v$; i.e., FWHM), and calculated pseudo-equivalent width (pEW). In Figure~\ref{fig:all_vsi} we show the velocity evolution of the two components compared to the \vsi\ evolution of other \sneia\ from \citet{benetti05}. The HVF component shows a strong velocity gradient ($\dot{v}_{\rm Si}=$~353~km\,s$^{-1}$\,day$^{-1}$) but at velocities much higher than those seen in most \sneia, even at early times. The photospheric component, on the other hand, remains virtually constant in velocity even to late times (see \S~\ref{sec:late_vsi}), except for tentative evidence for higher velocity at the earliest epochs. However, we caution that the photospheric component is much weaker than the HVF component at those phases, so the velocity is more uncertain.

\begin{table}
\begin{center}
\caption{\siline\ Fit Results}
\label{tab:vsi_fits}
\begin{tabular}{rrrrrrr}
\hline
 & \multicolumn{3}{c}{HVF} & \multicolumn{3}{c}{Photospheric} \\
Phase  & $v$    & $\Delta v$ & pEW   & $v$    & $\Delta v$ & pEW \\
(days) & (km/s) & (km/s)     & (\AA) & (km/s) & (km/s)     & (\AA) \\
\hline
-14.51 & 22704 & 11829 & 164.2 & --    & --   & --   \\
-14.17 & 22233 & 11587 & 171.6 & --    & --   & --   \\
-12.66 & 21525 &  9046 & 125.4 & 13444 & 6589 & 24.3 \\
-12.53 & 21468 &  8898 & 120.2 & 13444 & 6294 & 23.0 \\
-11.51 & 21223 &  7665 &  82.0 & 13430 & 6473 & 30.5 \\
-10.05 & 20798 &  6315 &  43.5 & 13095 & 6547 & 38.0 \\
 -9.56 & 20374 &  6642 &  41.8 & 12685 & 6262 & 40.4 \\
 -9.35 & 20416 &  6326 &  37.6 & 12755 & 6431 & 43.0 \\
 -8.99 & 20275 &  6188 &  33.8 & 12595 & 6241 & 41.0 \\
 -8.47 & 20431 &  5767 &  29.5 & 12656 & 6347 & 46.8 \\
 -8.32 & 20327 &  5735 &  28.2 & 12571 & 6241 & 46.9 \\
 -7.97 & 20147 &  5503 &  22.6 & 12538 & 6041 & 46.2 \\
 -7.70 & 19902 &  6304 &  26.2 & 12057 & 6072 & 48.7 \\
 -7.54 & 20105 &  5735 &  24.5 & 12397 & 6167 & 51.3 \\
 -6.82 & 19501 &  5334 &  16.2 & 12284 & 5999 & 52.1 \\
 -6.67 & 19463 &  5545 &  16.2 & 11977 & 5978 & 53.6 \\
 -6.43 & 19817 &  4892 &  15.9 & 12251 & 6231 & 58.5 \\
 -5.66 & 19902 &  5651 &  16.1 & 12203 & 5841 & 55.5 \\
 -4.77 & 18713 &  5071 &  10.0 & 12104 & 5535 & 56.9 \\
 -4.70 & 18657 &  5345 &   9.8 & 11892 & 5619 & 58.3 \\
 -3.80 & 18397 &  5197 &   8.8 & 12048 & 5450 & 59.6 \\
 -3.70 & 18572 &  4892 &   5.5 & 11821 & 5398 & 58.4 \\
 -2.52 & 17242 &  4934 &   7.4 & 11944 & 5250 & 61.5 \\
 -2.42 & 18478 &  5693 &   8.0 & 12010 & 5377 & 63.9 \\
 -1.47 & 18006 &  4322 &   4.2 & 12057 & 5155 & 63.9 \\
 -0.37 & --    &  --   &    -- & 12034 & 4965 & 63.4 \\
 +0.34 & --    &  --   &    -- & 11821 & 4776 & 61.7 \\
 +0.70 & --    &  --   &    -- & 12180 & 4755 & 62.9 \\
 +1.22 & --    &  --   &    -- & 12175 & 4818 & 64.0 \\
 +1.71 & --    &  --   &    -- & 12095 & 4617 & 62.8 \\
 +3.18 & --    &  --   &    -- & 12156 & 4533 & 64.0 \\
 +4.47 & --    &  --   &    -- & 12180 & 4270 & 62.3 \\
 +5.42 & --    &  --   &    -- & 12109 & 4185 & 62.2 \\
 +6.38 & --    &  --   &    -- & 12123 & 4132 & 62.4 \\
 +7.52 & --    &  --   &    -- & 12227 & 3995 & 61.4 \\
 +8.44 & --    &  --   &    -- & 12104 & 3953 & 61.1 \\
 +9.20 & --    &  --   &    -- & 12118 & 3953 & 60.7 \\
 +9.64 & --    &  --   &    -- & 12071 & 3953 & 61.3 \\
+11.21 & --    &  --   &    -- & 12006 & --   & --   \\
+17.38 & --    &  --   &    -- & 11732 & --   & --   \\
+22.17 & --    &  --   &    -- & 11423 & --   & --   \\
+26.46 & --    &  --   &    -- & 11794 & --   & --   \\
+30.18 & --    &  --   &    -- & 11829 & --   & --   \\
+35.37 & --    &  --   &    -- & 11771 & --   & --   \\
+39.18 & --    &  --   &    -- & 11716 & --   & --   \\
\hline
\end{tabular}
\end{center}
\end{table}

\begin{figure}
\begin{center}
\includegraphics[width=0.45\textwidth]{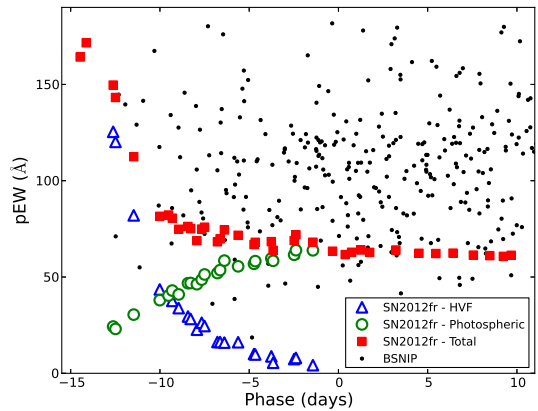}
\caption{Pseudo-equivalent width (pEW) of the \siline\ components in SN~2012fr as a function of phase. When both the HVF and photospheric components are clearly detected, they are shown as open blue triangles and open green circles, respectively. The total pEW of the \siline\ line is shown as filled red squares. For references, the BSNIP sample \citep{bsnip1, bsnip2} is show as small black points.}
\label{fig:pEW_evolution}
\end{center}
\end{figure}

The relative strength of the two components is most clearly captured by examining the absorption strength of each component as quantified by the pEW. This can be trivially calculated as the area of the normalized absorption profile. We show in Figure~\ref{fig:pEW_evolution} the pEW of the two fitted components from the earliest epoch ($-13$ days) to the latest epoch ($-1$ days) at which both features have a significant detection, and for the full \siline\ profile for all epochs before +10 days. As previously noted, the strength of the HVF fades very quickly while that of the photospheric component slowly rises, with the equality point occurring between $-10$ and $-9$ days. The total pEW of the \siline\ line declines until a few days before maximum light, when it remains nearly constant at around 65~\AA. As we discuss below (\S~\ref{sec:classification}), the pEW of this line is lower in SN~2012fr than in many other normal \sneia\ as measured in the Berkeley SN Ia Program (BSNIP) sample \citep{bsnip1, bsnip2}.

\subsection{Full Velocity Evolution of \siline}
\label{sec:late_vsi}
After the HVF \siline\ feature fades, the main photospheric component is well fit by a single Gaussian until about two weeks after maximum light. At that time, Fe lines to the red and blue of \siline\ begin to develop significant opacity and make it impossible to correctly determine the pseudo-continuum of the \siline\ line.

\begin{figure}
\begin{center}
\includegraphics[width=0.45\textwidth]{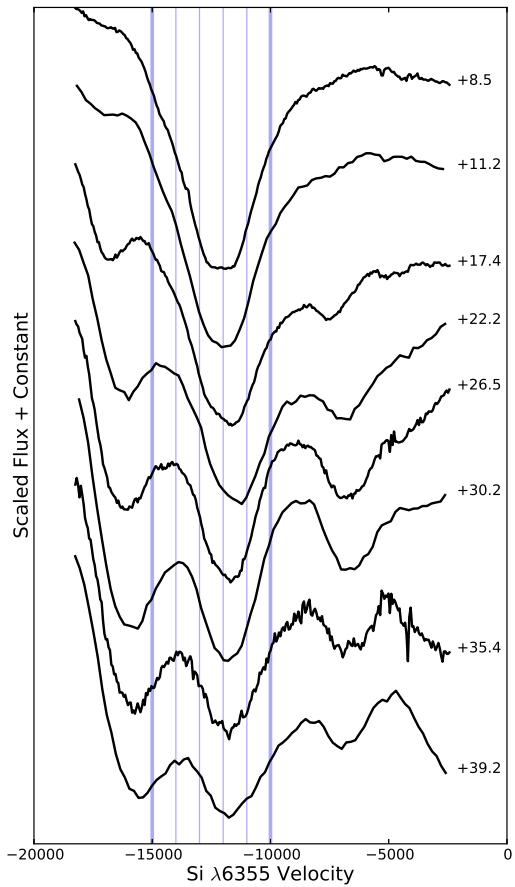}
\caption{Late-time evolution of the \siline\ line in SN~2012fr, shown in velocity space. The wavelengths corresponding with 10,000~km~s$^{-1}$ and 15,000~km~s$^{-1}$ are shown as thick blue lines, with 1,000~km~s$^{-1}$ intervals denoted by thin blue lines. The velocity plateau at late times appears around 11,800--12,000~km~s$^{-1}$.}
\label{fig:late_vsi}
\end{center}
\end{figure}

Thus, for the epochs +17 days and later, we fit the velocity minimum of \siline\ by fitting a simple Gaussian profile only in a region of width 30 \AA\ centered on the \siline\ minimum. Remarkably, the velocity of the line minimum remains nearly constant at about 11,800~km~s$^{-1}$ even out to phase +39 d. This is confirmed by visual inspection of the \siline\ region as plotted in Figure~\ref{fig:late_vsi}. We do note that at such late epochs, emission becomes increasingly important \citep[see, e.g.,][]{vanrossum12} in the line profiles, so it is possible that the flux minimum may not necessary trace the true $\tau=1$ surface.

We see once again that the narrow width of the photospheric lines provides an advantage in following the \siline\ velocity reliably to very late epochs. The Fe lines to the red and blue of this feature are clearly distinguished in SN~2012fr, enabling an accurate isolation of \siline\ and a reliable measurement of its velocity. In most other \sneia, the broader line widths result in a blend of the \siline\ with its neighboring Fe lines, making velocity measurements difficult at late epochs. We will return to this point in \S~\ref{sec:silicon_velocities}.

\section{Additional Atomic Species in SN~2012fr}
\label{sec:other_lines}
The narrow velocity width of the photospheric \siline\ line noted in \S~\ref{sec:si_line} holds true for nearly all absorption features in the optical spectra of SN~2012fr starting about a week after maximum light. This provides a unique advantage in line identifications in the SN spectra, as blending of neighboring lines is less pronounced in SN~2012fr than in many other \sneia.

In this section we focus on four element groups of particular interest. We begin by inspecting narrow \naid\ absorption in \S~\ref{sec:sodium}, and show that SN~2012fr shows no detectable absorption in this line. We briefly present in \S~\ref{sec:carbon} our search for signatures of unburned C, which showed no clear detection. In \S~\ref{sec:caii} we inspect the \canir, which exhibits behavior similar to that of \siline, and then briefly examine the more complex \cahk\ line at maximum light. Finally, in \S~\ref{sec:fe_group}, we examine the velocities of Fe-group elements in SN~2012fr.

\subsection{\naid\ Narrow Absorption}
\label{sec:sodium}


Narrow absorption in the \naid\ line in \snia\ spectra is commonly used to quantify the amount of foreground dust that reddens a \snia\ (dust is associated with the detected gas). \naid\ absorption at the redshift of the SN host galaxy presumably arises from either foreground interstellar gas \citep[e.g.,][]{poznanski11, poznanski12} or very nearby circumstellar material shed from the SN progenitor system prior to explosion \citep[][]{patat07, simon09, sternberg11, dilday12}.

\begin{figure}
\begin{center}
\includegraphics[width=0.45\textwidth]{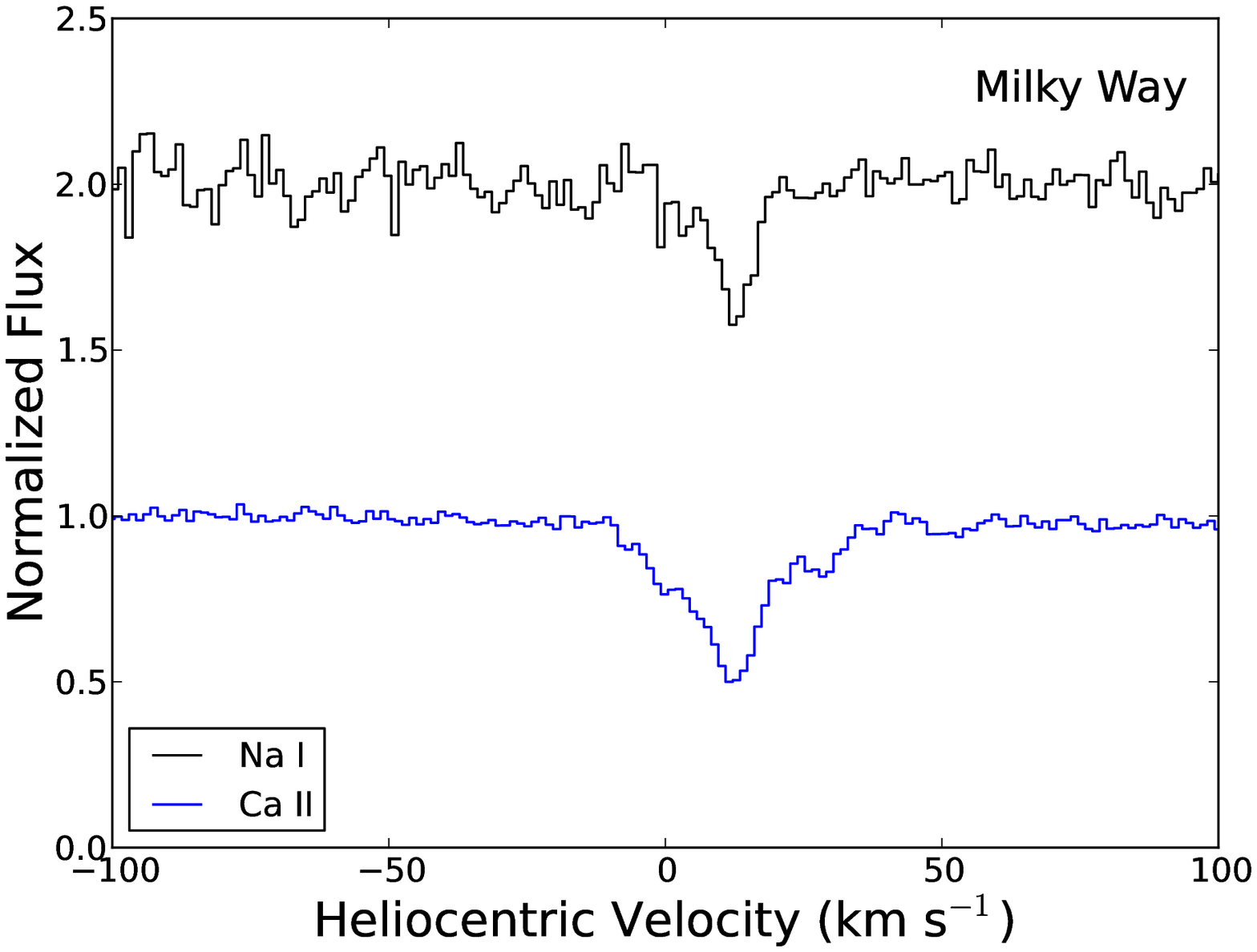}
\includegraphics[width=0.45\textwidth]{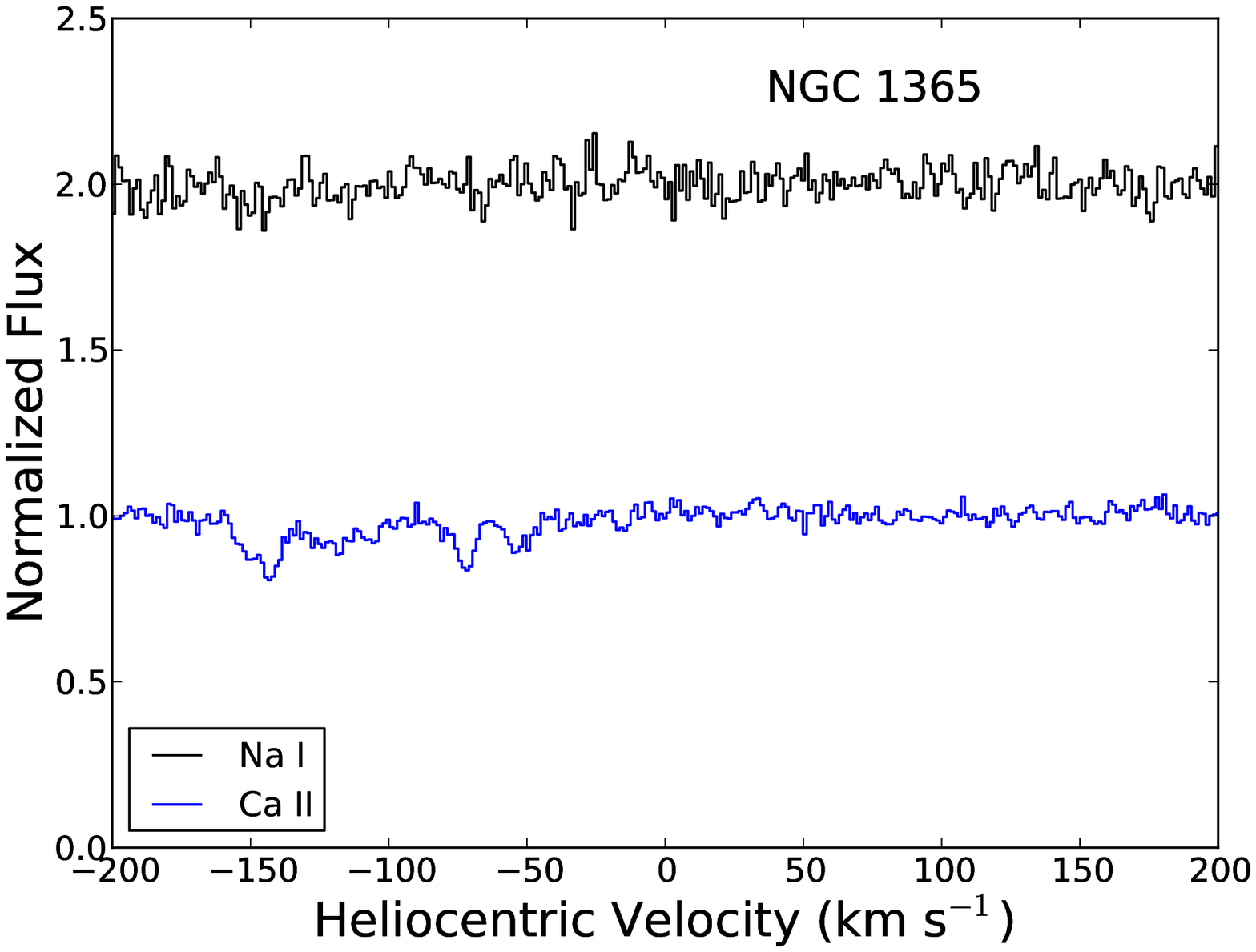}
\caption{Keck HIRES observations of SN~2012fr. Top: Narrow \naid\ and \CaII\ H\&K at rest velocity, arising from Milky Way gas. Bottom: Narrow \naid\ and \CaII\ H\&K at the recession velocity of NGC~1365.}
\label{fig:narrow_sodium}
\end{center}
\end{figure}

SN~2012fr shows no detectable narrow \naid\ absorption in the HIRES spectrum taken at phase $-13.6$ days. In Figure~\ref{fig:narrow_sodium}  we show the regions of the HIRES spectrum corresponding to the \naid\ line as well as the \CaII\ H\&K lines, both at the redshift of the host galaxy NGC~1365 \citep[$v=1636$~km~s$^{-1}$;][]{zNGC1365} and at zero redshift. Measurements of the line detections, or $3\sigma$ upper limits, are presented in Table~\ref{tab:hires_results}.

\begin{table}
\begin{center}
\caption{HIRES Absorption Equivalent Widths}
\label{tab:hires_results}
\begin{tabular}{lrr}
\hline
Line             & Milky Way & NGC~1365 \\
                 & (m\AA)    & (m\AA) \\
\hline
\CaII\ $\lambda$3934.777  & $129.3 \pm  3.6$ & $107.3 \pm 7.0$ \\
\CaII\ $\lambda$3969.591  & $ 66.0 \pm  4.0$ & $ 57.3 \pm 6.0$ \\
\NaI\  $\lambda$5891.5833 & $ 82.9 \pm 13.8$ & $<42.0$~($3\sigma$) \\
\NaI\  $\lambda$5897.5581 & $ 35.4 \pm 13.4$ & $<40.8$~($3\sigma$) \\
\hline
\end{tabular}
\end{center}
\end{table}

Narrow absorption features from Milky Way gas are clearly detected at $2.6\sigma$ and $6.0\sigma$ in the D1 and D2 lines of \NaI, and at very high significance ($>10\sigma$) in the H\&K lines of \CaII. Using the empirical scaling relations of \citet{poznanski12} to convert \naid\ absorption into the reddening $E(B-V)$, the measured D1 and D2 absorption strengths imply reddenings of $E(B-V) = 0.0186$ and $E(B-V) = 0.0213$ mag, respectively. These are in excellent agreement with the measured value of $E(B-V) = 0.018$ mag from \citet{sf11}.


The only possible absorption features at the redshift of NGC~1365 appear in the \CaII\ H\&K lines at $v \approx -100$~km~s$^{-1}$ from the rest redshift of NGC~1365. Given that these features lack corresponding ones in \naid, in addition to the facts that NGC~1365 is a nearly face-on barred spiral and SN~2012fr is very close to the center of the galaxy and thus well away from the dusty spiral arms, it appears unlikely that this feature near \CaII\ H\&K is truly caused by interstellar gas. Thus, we detect no significant narrow absorption features in SN~2012fr. Given the traditional correlation of these narrow absorption features with reddening by foreground dust, our observations are consistent with SN~2012fr having no obscuration by foreground dust. The strongest constraint arises from the D2 line of \NaI, which places a $3\sigma$ upper limit of $E(B-V) < 0.015$ mag \citep[again using][]{poznanski12} for the reddening of SN~2012fr from within NGC~1365.

\subsection{\CII}
\label{sec:carbon}
We comment briefly in this section on the search for unburned C features in spectra of SN~2012fr.  Such signatures typically manifest themselves as weak \CII\ absorption lines in optical spectra of \sneia, and have been of particular interest in recent years \citep{rcthomas11, parrent11, folatelli12, bsnip4, blondin12}. The strongest C feature in the optical is typically the \CII\ $\lambda6580$ line, and the slightly weaker $\lambda7234$, $\lambda4745$, and $\lambda4267$ lines of \CII\ are also sometimes visible \citep[e.g.,][]{mazzali01b, rcthomas07}.

The redshift of NGC~1365 places the likely location of the blueshifted absorption minimum of \CII\ $\lambda$6580 coincident with a weak telluric absorption feature at $\lambda=6280$~\AA\ (corresponding to $v \approx$ 13,700~km~s$^{-1}$ for \CII\ $\lambda6580$). This telluric line is of comparable strength to the typical \CII\ absorption, but it was entirely or largely removed during the reduction process. We cannot identify any obvious signature of \CII\ $\lambda$6580 absorption at its position; nor do we detect the other major optical \CII\ lines even at early epochs.

\begin{figure}
\begin{center}
\includegraphics[width=0.45\textwidth]{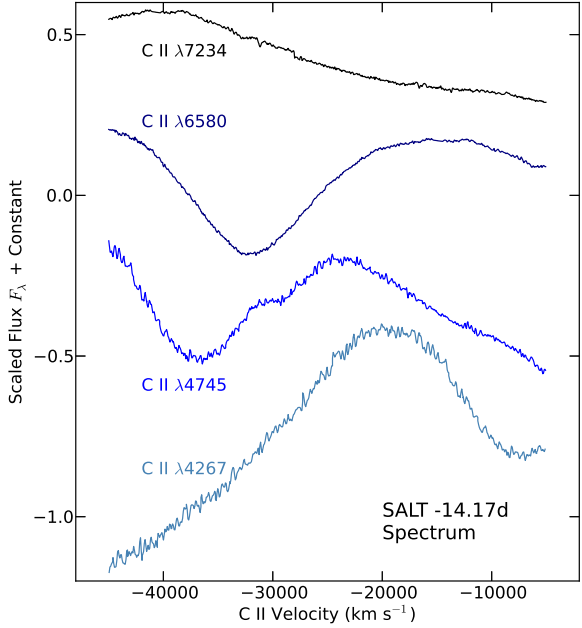}
\caption{Sections of the $-14.17$ day SALT spectrum of SN~2012fr corresponding to the strongest typical \CII\ features in \sneia, plotted in terms of velocity of the respective lines.}
\label{fig:carbon}
\end{center}
\end{figure}

This is demonstrated directly in Figure~\ref{fig:carbon}, where we plot regions around the four \CII\ lines in velocity space for the $-14.17$ day spectrum from SALT. No clear signature of \CII\ seems visible in any of these lines, even for \CII\ $\lambda$6580 after removal of the telluric feature. \CI\ features in the NIR may provide better detection of unburned material in SN~2012fr \citep[see, e.g., SN~2011fe in][]{hsiao11fe}, and will be investigated with detailed spectroscopic fitting in Paper~III (Hsiao et al. 2013, in prep.).

\subsection{\CaII}
\label{sec:caii}
\subsubsection{\CaII\ Infrared Triplet}
\label{sec:ca_nir_triplet}
The \canir\ in SN~2012fr begins at very high velocities two weeks before maximum light, with the blue edge of the absorption in the first spectrum reaching relativistic velocities of nearly 50,000 km~s$^{-1}$ ($v \approx 0.17c$). The line complex gradually recedes in velocity and exhibits complex structure roughly a week before maximum light, and by two weeks after maximum it appears dominated by a single narrow component.

The complex structure of the \canir\ appears to be indicative of multiple components in velocity space, similar to that seen in the \siline\ line (see \S~\ref{sec:hvf_si}). Modeling this line multiplet is more complex than the simple two-component Gaussian fits employed for \siline. Instead, each component of the \canir\ in velocity space must be modeled as a triplet of Gaussian profiles with common velocity width, separation in velocity space as dictated by the line rest wavelengths, and with relative absorption depths appropriately constrained.

\begin{figure}
\begin{center}
\includegraphics[width=0.45\textwidth]{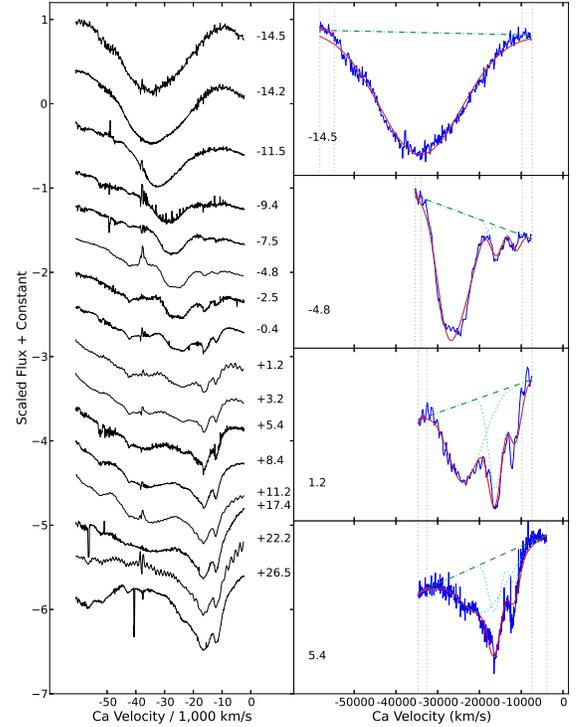}
\caption{Same as Figure~\ref{fig:twopeak_si_fits} but for the \canir. Here velocities are plotted with respect to the reddest line in the \canir\ at 8662~\AA.}
\label{fig:twopeak_ca_fits}
\end{center}
\end{figure}

For epochs where spectral coverage extended to sufficiently red wavelengths to cover the \canir, we fit the absorption profile as a two-component model after normalizing to a fitted pseudo-continuum. Each absorption component is described by a central velocity, velocity width, and absorption depth (here we set relative absorption depths of the triplet lines to be equal, assuming the optically thick regime). As with \siline, the only constraint applied here was to force the two components to occupy different regions of velocity space split at 14,000 km~s$^{-1}$. We show the spectral evolution of the \canir\ as well as some representative profile fits in Figure~\ref{fig:twopeak_ca_fits}.
As with the \siline\ fits, the pseudo-continuum shape begins to be poorly represented by a simple linear fit at late times. Thus we employ a similar Gaussian line minimum fitting technique as that for \siline\ at epochs after +8 days, here measuring the minimum of the cleanly separated 8662~\AA\ line.

\begin{figure}
\begin{center}
\includegraphics[width=0.45\textwidth]{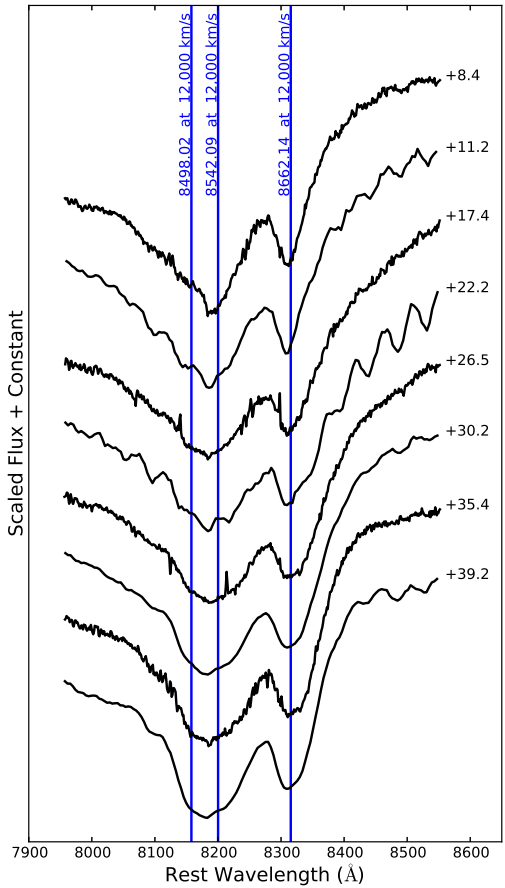}
\caption{Late-epoch evolution of the \canir\ in SN~2012fr, showing the narrow velocity width and constant velocity of the three lines comprising the triplet. For reference, we mark the wavelengths of the three triplet lines at a velocity of $v =$ 12,000~km~s$^{-1}$, showing that the reddest line of the triplet at 8662~\AA\ is cleanly resolved from the bluer two lines.}
\label{fig:late_vca}
\end{center}
\end{figure}

We report results in Table~\ref{tab:canir_fits}, with fit parameters labeled similarly as in Table~\ref{tab:vsi_fits}. 
The high-velocity component consistently exhibits a broad velocity width ($>5000$~km~s$^{-1}$), producing a broad component where all lines in the triplet are blended.
The low-velocity component exhibits the same narrow velocity width as observed in the \siline\ line, making it possible to distinguish the 8662~\AA\ line from the blended 8498~\AA\ and 8542~\AA\ lines. This is illustrated directly in Figure~\ref{fig:late_vca}, where we show the \canir\ evolution at late times and demonstrate the ability to both resolve the triplet lines and observe their consistent velocity at $v \approx$ 12,000~km~s$^{-1}$.

\begin{table}
\begin{center}
\caption{\canir\ Fit Results}
\label{tab:canir_fits}
\begin{tabular}{rrrrrrr}
\hline
 & \multicolumn{3}{c}{HVF} & \multicolumn{3}{c}{Photospheric} \\
Phase  & $v$    & $\Delta v$ & pEW & $v$    & $\Delta v$ & pEW \\
(days) & (km/s) & (km/s)     & (\AA) & (km/s) & (km/s)     & (\AA) \\
\hline
-14.51 & 31087 & 21853 & 615.5 & --    & --   & --   \\
-11.51 & 28640 & 15311 & 364.8 & --    & --   & --   \\
 -9.35 & 25571 &  9581 & 146.9 & --    & --   & --   \\
 -7.54 & 24126 &  5507 &  94.5 & 11221 & 1629 &  4.1 \\
 -4.77 & 23043 &  5808 &  99.7 & 11259 & 2407 & 13.2 \\
 -2.52 & 21994 &  6226 &  96.1 & 11219 & 2697 & 29.9 \\
 -0.37 & 20989 &  6662 &  90.0 & 11357 & 3151 & 49.5 \\
 +1.22 & 20131 &  6935 &  82.9 & 11461 & 3187 & 60.3 \\
 +3.18 & 19133 &  7368 &  78.2 & 11564 & 3207 & 70.8 \\
 +5.42 & 15913 & 10582 & 121.5 & 11690 & 2769 & 52.9 \\
 +8.44 & --    &  --   &    -- & 12206 & --   & --   \\
+11.21 & --    &  --   &    -- & 12240 & --   & --   \\
+17.38 & --    &  --   &    -- & 12092 & --   & --   \\
+22.17 & --    &  --   &    -- & 12121 & --   & --   \\
+26.46 & --    &  --   &    -- & 11976 & --   & --   \\
+30.18 & --    &  --   &    -- & 12082 & --   & --   \\
+35.37 & --    &  --   &    -- & 11937 & --   & --   \\
+39.18 & --    &  --   &    -- & 12052 & --   & --   \\
\hline
\end{tabular}
\end{center}
\end{table}

\begin{figure}
\begin{center}
\includegraphics[width=0.45\textwidth]{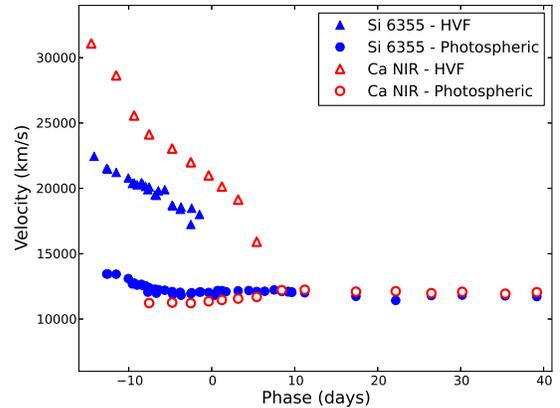}
\caption{Velocity evolution of the high-velocity (triangles) and photospheric (circles) components of the \canir\ (open red symbols) in SN~2012fr compared to that of the \siline\ line (filled blue symbols).}
\label{fig:all_vca}
\end{center}
\end{figure}

In Figure~\ref{fig:all_vca} we plot the velocity evolution of the fitted \canir\ components compared to the analogous components in the \siline\ line (\S~\ref{sec:si_line}). The HVF \canir\ component is consistently at higher velocities than the \siline\ HVF, has a steeper velocity gradient ($\dot{v}_{\rm Ca} = 686$~km\,s$^{-1}$\,day$^{-1}$), and is visible to later epochs than the \siline\ HVF \citep[we note that these characteristics were also observed in the \canir\ feature of SN~2009ig;][]{marion13}. In SN~2012fr the late-time velocity of the \canir\ plateaus at $v \approx$ 12,000~km~s$^{-1}$, consistent with the velocity plateau observed in the \siline\ line.

Finally, we comment on the dependence of our results on the assumption of optical thickness in the \canir. We repeated the above fits under the assumption of the optically thin regime, where the relative absorption depths of the triplet lines are proportional to their Einstein $B$ (absorption) values ($1.71\times 10^9\mathrm{cm}^2\,\mathrm{s}^{-1}\,\mathrm{erg}^{-1}$ for $\lambda$8498, $1.03\times 10^{10}\mathrm{cm}^2\,\mathrm{s}^{-1}\,\mathrm{erg}^{-1}$ for $\lambda$8542, $8.66\times 10^9\mathrm{cm}^2\,\mathrm{s}^{-1}\,\mathrm{erg}^{-1}$ for $\lambda$8662; \citealt{wiese69}) and the relevant statistical weights, and found that the qualitative behavior of the velocity evolution was consistent with that found in our fiducial fits. The velocities of the HVF \canir\ component were consistently higher due to the shift in weighted mean wavelength of the triplet, but the late-time velocity was still consistent with the $v \approx$ 12,000~km~s$^{-1}$ velocity plateau due to the 8662~\AA\ line being distinguishable from the bluer lines in the triplet. Interestingly, the photospheric component appears to be better fit (and yields a velocity consistent with the velocity plateau) in the optically thin assumption at early epochs, and then transitions to being optically thick around the time when the HVF component fades. The true absorption strengths of the lines in the \canir\ likely fall somewhere between the optically thin and optically thick regimes, but we have confirmed that our results are consistent in both extreme cases.

\subsubsection{\cahk}
The \cahk\ doublet in \sneia\ is a line complex of keen interest, as its behavior at maximum light may be an indicator of intrinsic \snia\ color \citep{foley11, chotard11, blondin12, foley12}. Additionally, recent work by \citet{maguire12} showed that the \cahk\ velocity of the mean rapidly declining (low ``stretch'') \snia\ spectrum at maximum light is lower than that of the mean slowly declining (high ``stretch'') \snia\ spectrum. Thus, the \cahk\ line seems promising for helping to unravel \snia\ diversity.

Interpreting \CaII\ velocities with this line complex is difficult, however, due to the presence of the nearby \siblue\ line \citep[see the thorough discussion in][]{foley12}, as well as the complex underlying pseudo-continuum. We therefore chose to examine this line complex only at maximum light for SN~2012fr. We employed a fitting procedure similar to that of the \canir, but with fit parameters informed and tightly constrained by results of the \canir\ and \siline\ fits. We model absorption in the \cahk\ line with multiple velocity components, where each component is a doublet profile with relative depths set to unity (i.e., the optically thick regime). In addition to the HVF and photospheric components of the \cahk\ line, we add a single Gaussian absorption profile to model \siblue. We note that fits assuming the optically thin regime (with line depths proportional to the Einstein $B$ values of $4.50\times10^{10}\mathrm{cm}^2\,\mathrm{s}^{-1}\,\mathrm{erg}^{-1}$ for $\lambda$3934 and $2.20\times10^{10}\mathrm{cm}^2\,\mathrm{s}^{-1}\,\mathrm{erg}^{-1}$ for $\lambda$3968; \citealt{wiese69}) produced poor fits to the \cahk\ line, especially the photospheric component.

\begin{figure}
\begin{center}
\includegraphics[width=0.45\textwidth]{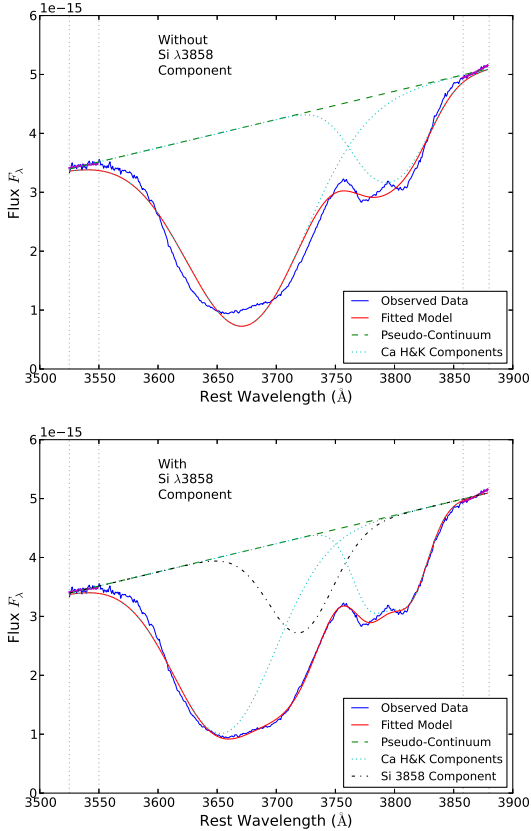}
\caption{Fit to the \cahk\ line profile in the maximum-light (\wifes\ Nov. 11.67) spectrum of SN~2012fr. Line colors and styles are the same as in Figure~\ref{fig:twopeak_si_fits}, but with the \siblue\ line profile plotted as the dashed-dotted black line. The top panel shows a fit without the \siblue\ line, while the bottom panel shows a fit including this line.}
\label{fig:cahk_bmax}
\end{center}
\end{figure}

To enforce consistency with the other lines measured for Ca and Si, we constrained the velocity center and velocity widths of the two \cahk\ components to be within 20\% of their values fitted for the \canir, but left the absorption depths as free parameters. Similarly, we forced the velocity width and center of the \siblue\ line to be within 20\% of the \siline\ line. To investigate the importance of the \siblue\ line in this complex, we performed fits both with and without the \siblue\ line included, and our best fits for each case are shown in Figure~\ref{fig:cahk_bmax}.

\begin{table*}
\begin{center}
\caption{\cahk\ Fit Results}
\label{tab:cahk_fits}
\begin{tabular}{lrrrrrrrrr}
\hline
 & \multicolumn{3}{c}{Ca HVF} & \multicolumn{3}{c}{Ca Photospheric} & \multicolumn{3}{c}{Si Photospheric} \\
  & $v$    & $\Delta v$ & pEW & $v$    & $\Delta v$ & pEW & $v$    & $\Delta v$ & pEW \\
  & (km\,s$^{-1}$) & (km\,s$^{-1}$)     & (\AA) & (km\,s$^{-1}$) & (km\,s$^{-1}$)     & (\AA) & (km\,s$^{-1}$) & (km\,s$^{-1}$)     & (\AA) \\
\hline
Ca NIR + \siline\  & 20989 & 6662 &  96.1 & 11357 & 3151 & 49.5 & 12034 & 4965 &   -- \\
Ca H\&K + \siblue\ & 22545 & 6633 &  79.9 & 11775 & 2675 & 23.6 & 10749 & 4707 & 25.4 \\
Ca H\&K only       & 21216 & 7994 & 103.3 & 11787 & 3665 & 23.8 &    -- &   -- &   -- \\
\hline
\end{tabular}
\end{center}
\end{table*}

In Table~\ref{tab:cahk_fits} we summarize the main results of our \cahk\ line fits, for the cases with and without \siblue, as well as the \canir\ and \siline\ results for reference. Both \cahk\ fits yield Ca and Si velocities within 1000~km~s$^{-1}$ of their red counterparts, but the fit with \siblue\ included shows a more favorable ratio of pEW for the photospheric and HVF Ca components, as well as a much better fit to the overall line profile. Our fits indicate that both \cahk\ and \siblue\ are needed to explain the absorption profile of the \cahk\ line complex in SN~2012fr, with the HV \cahk\ being dominant over the \siblue.


We note, however, that the quantitative details of these results depend on how tightly we constrain the velocity center and width of the HV \cahk\ component. If we loosened the constraints to be within 30\% of the \canir\ velocity and width, it changes the pEW of the HV \cahk\ and \siblue\ lines to be 68~\AA\ and 35~\AA, respectively, from 80~\AA\ and 24~\AA\ when constrained to 20\%. This is because the high velocities of the HV \cahk\ component place it nearly coincident with the wavelength of the lower velocity \siblue\ line. We therefore cannot say conclusively what the absorption ratio of these two lines is in this line complex, but we did find consistently that both were needed to adequately fit the absorption profile.

We found here that decoupling the \siblue\ line from the \cahk\ line is a nontrivial procedure. While it is difficult to derive precise quantitative results, two general qualitative results are clear. The first is that both the \cahk\ line and the \siblue\ line are operative in this line complex in SN~2012fr, and that high-velocity \cahk\ can be nearly degenerate with lower velocity \siblue. The second conclusion for SN~2012fr is that regardless of the details of how much the line velocities and velocity widths are allowed to vary, high-velocity \cahk\ appears to be the dominant contributor to absorption in this line region.

\subsection{Fe-Group Elements}
\label{sec:fe_group}
\begin{figure}
\begin{center}
\includegraphics[width=0.45\textwidth]{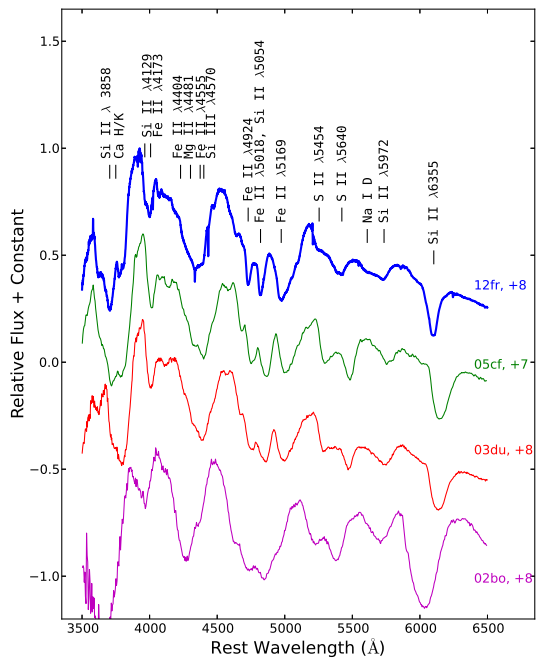}
\caption{Line identifications in SN~2012fr (blue) for the +8 day spectrum, compared to the spectra of other \sneia\ at a similar epoch. SN~2005cf \citep{wang05cf}, SN~2003du \citep{stanishev03du}, and SN~2002bo \citep{benetti04} are shown as the green, red, and magenta spectra, respectively.}
\label{fig:line_iden}
\end{center}
\end{figure}

The relatively narrow line widths of SN~2012fr are useful for identifying absorption features which are typically blended in other \sneia, and are particularly advantageous for identifying Fe lines. In Figure~\ref{fig:line_iden} we illustrate this principle with the +8 day spectrum of SN~2012fr compared to two other ``normal'' \sneia, SN~2005cf \citep{wang05cf} and SN~2003du \citep{stanishev03du}, as well as to the broad-lined SN~2002bo \citep{benetti04}. Of particular note is the line complex at $\sim 4700$~\AA, comprising several \FeII\ lines and the \SiII\ $\lambda$5054 line, which shows cleaner separation in SN~2012fr than the other \sneia. 

\begin{figure}
\begin{center}
\includegraphics[width=0.45\textwidth]{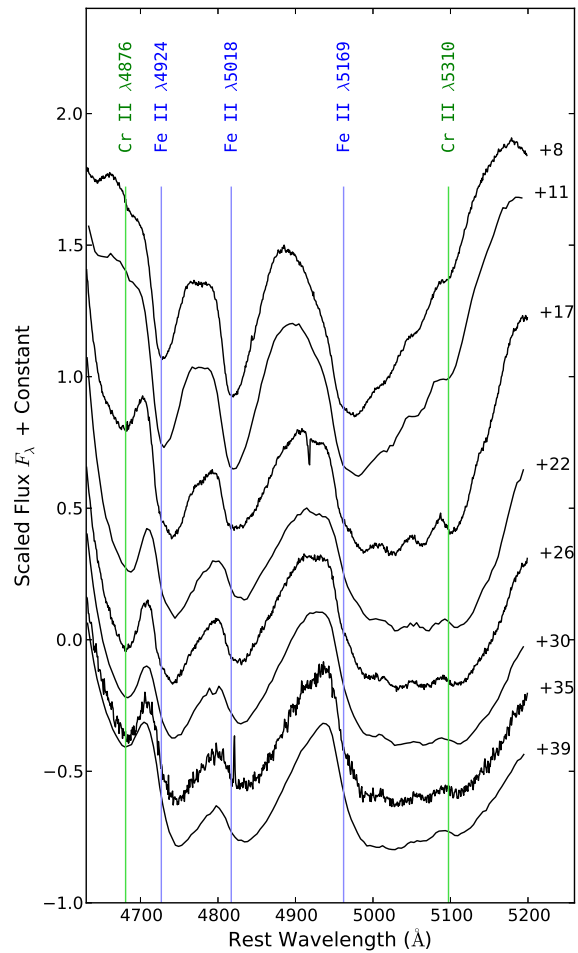}
\caption{Post-maximum spectra of SN~2012fr highlighting a selected region having significant features of Fe-group elements. Vertical lines correspond to wavelengths of labeled features at $v =$ 12,000~km~s$^{-1}$, the measured velocity of \FeII\ features in the +8 day spectrum.}
\label{fig:late_fegroup}
\end{center}
\end{figure}

In the +8 day spectrum, three major \FeII\ lines ($\lambda$4924, $\lambda$5018, and $\lambda$5169) all have velocity minima consistent with the velocity plateau ($v \approx$ 12,000~km~s$^{-1}$) identified in \siline. We illustrate this in Figure~\ref{fig:late_fegroup}, where we plot the post-maximum spectra of SN~2012fr in the relevant wavelength region and highlight the wavelengths associated with each of these lines at $v=$12,000~km~s$^{-1}$. After this epoch, the observed velocity minima of these Fe lines decrease with time.
We also tentatively identify two features which may be associated with \CrII\ lines ($\lambda$4876 and $\lambda$5310), though their velocity evolution is more difficult to follow due to the strongly evolving shape of the underlying pseudo-continuum.
We note here that these lines which we attribute to Fe-group elements show no clear signature of high-velocity features in the early-time spectra of SN~2012fr \citep[some Fe HVFs have been possibly identified in several \sneia; e.g.,][]{branch04, mazzali05a, marion13}, though further modeling of the spectra may reveal more subtle insights.

The velocity behavior of the Fe group lines indicates that the $\tau=1$ surface of the photosphere is at $v \approx$ 12,000~km~s$^{-1}$ one week after maximum brightness, and recedes inward after that time. From this we can conclude two important results: (i) the Si and Ca velocity plateaus imply that the layer of intermediate-mass elements (IMEs) is unlikely to extend deeper than $v \approx$ 12,000~km~s$^{-1}$ since the Fe-group elements are detected at lower velocities at the same epochs; and (ii) the layer of Fe-group elements extends deeper in the ejecta than the IMEs, indicating a likely stratification of the ejecta.

While these preliminary line identifications are interesting, a full accounting of all elements and their velocities would require a spectrum-synthesis fit \citep[see, e.g.,][]{synapps} such as that undertaken by \citet{parrent12} for SN~2011fe \citep{nugent11,li11fe}, or a modeling of abundance stratification in the ejecta such as that undertaken by \citet{stehle05}. Analyses of this magnitude are beyond the intended scope of this paper, but we believe that the spectra released here will be invaluable for such efforts, and we strongly encourage future modeling of the SN~2012fr spectra.

\section{Spectroscopic Subclassification of SN~2012fr}
\label{sec:classification}
In recent years much effort has been focused on categorizing the observed diversity of \sneia\ by means of quantitative metrics measured from their optical spectra. In this section we will examine SN~2012fr in the context of these classification schemes. The spectral indicator values of SN~2012fr employed for that purpose are presented in Table~\ref{tab:spec_inds}. These include several quantities (pEW(5972), pEW(6355), $v_{\rm Si}$) calculated from the maximum-light spectrum (the Nov. 11.67 \wifes\ spectrum), the velocity gradient of the \siline\ line $\dot{v}_{\rm Si}$ (measured from the absolute decline between phases 0 and +10 days), and the light-curve decline $\Delta m_{15}(B)$ (from Paper II). We note that the pEW values used here are calculated from direct integration of the line profile and differ insignificantly from the Gaussian area reported in \S~\ref{sec:hvf_si}.

\begin{table}
\begin{center}
\caption{Spectral Indicators for SN~2012fr}
\label{tab:spec_inds}
\begin{tabular}{lll}
\hline
Quantity & Value & Unit \\
\hline
$\Delta m_{15}(B)$ & $0.80 \pm 0.01$ & mag \\
pEW(5972)       & $3.9 \pm 5.0$   & \AA\ \\
pEW(6355)       & $66.5 \pm 15.5$ & \AA\ \\
$v_{\rm Si}$ (max)    & $12037 \pm 200$ & km\,s$^{-1}$ \\
$\dot{v}_{\rm Si}$    & $0.3 \pm 10.0$  & km\,s$^{-1}$\,day$^{-1}$ \\
\hline
\end{tabular}
\end{center}
\end{table}

\begin{figure*}
\begin{center}
\includegraphics[width=0.90\textwidth]{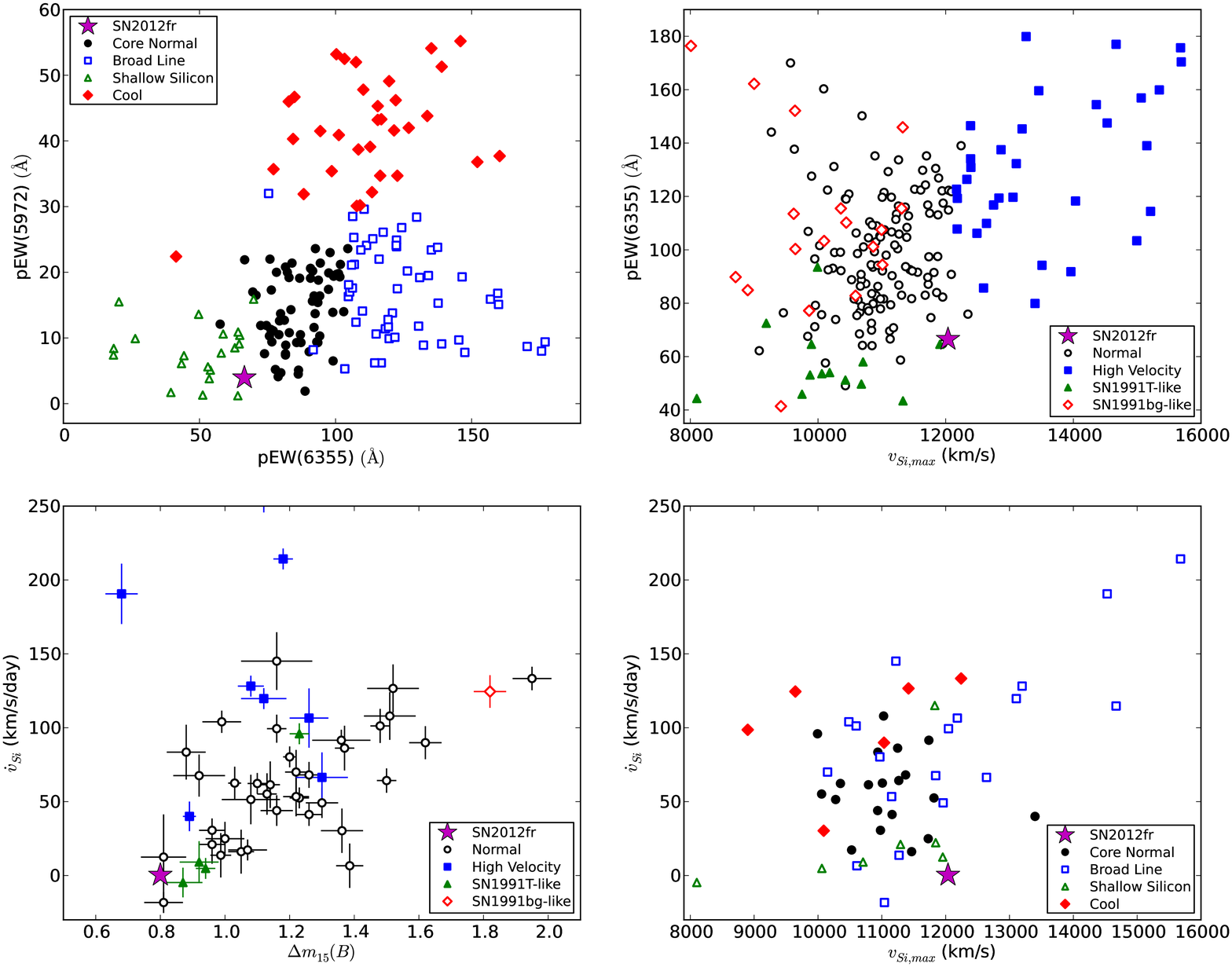}
\caption{Spectral indicators from SN~2012fr compared to those of other \sneia\ as measured by \citet{blondin12} and \citet{bsnip2}. Top Left: pEW(5972) vs. pEW(6355) at maximum light with spectroscopic subclasses as defined by \citet{branch09}. Top Right: pEW(6355) vs. $v_{\rm Si}$ at maximum light with velocity-based subclasses as defined by \citet{wang09}. Bottom Left: Velocity gradient of the \siline\ line $\dot{v}_{\rm Si}$, as measured from the absolute decline between phases 0 and +10 days, vs. light-curve decline $\Delta m_{15}(B)$, as previously inspected by \citet{benetti05}. Bottom Right: Velocity gradient vs. $v_{\rm Si}$ at maximum light.}
\label{fig:spec_inds}
\end{center}
\end{figure*}

\citet{branch09} proposed that \sneia\ can be split into four broad categories based on their location in the parameter space defined by the pEW of the \SiII\ $\lambda$5972 and $\lambda$6355 lines. We show SN~2012fr on this ``Branch diagram'' in the upper-left panel of Figure~\ref{fig:spec_inds}, along with a set of \sneia\ combining the samples of \citet{blondin12} and \citet{bsnip2}. SN~2012fr falls on the boundary between the ``shallow silicon'' class and the ``core normal'' class. As previously noted by both \citet{bsnip2} and \citet{blondin12}, these Branch classes do not represent disjoint samples with distinct features, but instead regions within a continuum of \snia\ characteristics. SN~2012fr appears to be a clear example of a transition-like event that bridges the gap between two subclasses.

\citet{wang09} showed that a subset of \sneia\ display high velocities in the \siline\ line, and this ``HV'' subset exhibits different color behavior than \sneia\ with normal velocities. In the upper-right panel of Figure~\ref{fig:spec_inds} we show SN~2012fr on the ``Wang diagram'' which plots pEW(6355) vs. $v_{\rm Si}$ at maximum light, as well as the same comparison sample of \sneia\ from the Branch diagram. Once again, SN~2012fr has a $v_{\rm Si}$ value which is just slightly above the established boundary separating HV from normal \sneia, marking it as a transition-like event between velocity classes. Its location in this diagram is noteworthy because it exhibits a lower pEW(6355) than any of the other HV \sneia, and its velocity is higher than that of other \sneia\ having weak \siline\ absorption.

\citet{benetti05} examined subclasses of \sneia\ based on the velocity evolution of the \siline\ line. They found that rapidly declining \sneia\ tended to have consistently high velocity gradients (dubbed the ``faint'' subclass), while slowly declining \sneia\ appeared to occur in two classes with either high or low velocity gradients (``HVG'' and ``LVG,'' respectively). In the bottom-left panel of Figure~\ref{fig:spec_inds}, we show SN~2012fr on this ``Benetti diagram'' along with the subset of \sneia\ from the other panels with sufficient data to measure a velocity gradient. SN~2012fr clearly resides in the LVG region of this diagram, as expected due to the observed \siline\ velocity plateau (see \S~\ref{sec:late_vsi}).

Finally, in the bottom-right panel of Figure~\ref{fig:spec_inds} we show the velocity gradient vs. velocity at maximum light. As has been previously noted, the HV objects appear to show a correlation between their velocity gradient and velocity at maximum light. SN~2012fr has a lower velocity gradient than any of the HV members, and instead appears to reside at the edge of the cloud of points populated by normal-velocity \sneia.

SN~2012fr exhibits much of the spectroscopic and photometric behavior of the more luminous \sneia\ such as SN~1991T \citep{phillips91T, filippenko91T} or SN~1999aa \citep{li01,garavini99aa}. It has a slow light-curve decline rate, relatively shallow \SiII\ absorption at maximum light, and a very low velocity gradient in the \siline\ line. However, two key features of SN~2012fr are not observed in SN 1999aa-like or SN 1991T-like \sneia: the high velocity of the \siline\ line (both at maximum light in the photospheric component and in the early HVF component), and strong absorption in the \siline\ line at early epochs (phase about $-10$ days). 

\begin{figure*}
\begin{center}
\includegraphics[width=0.90\textwidth]{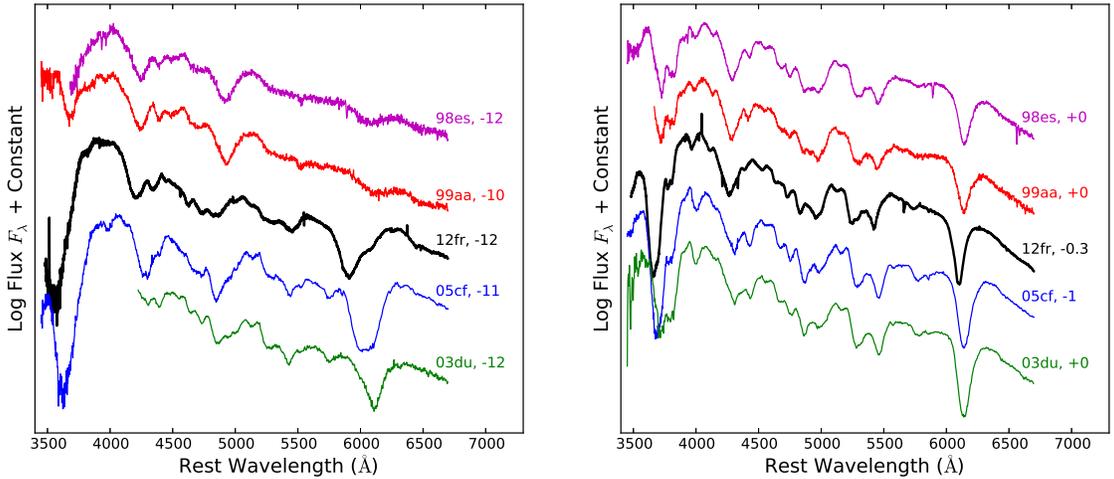}
\caption{Spectrum of SN~2012fr compared to other \sneia\ at $\sim 12$ days before maximum light (left) and at maximum light (right). Comparison \sneia\ are the SN~1991T-like SN~1998es \citep[from][]{blondin12}, SN~1999aa \citep[from][]{blondin12}, and the normal \sneia\ SN~2005cf \citep{wang05cf} and SN~2003du \citep{stanishev03du}.}
\label{fig:99aa_comparison}
\end{center}
\end{figure*}

In Figure~\ref{fig:99aa_comparison} we compare the spectra of SN~2012fr at $-12$ days and at maximum light to comparable spectra from the SN 1991T-like SN~1998es \citep[from][]{blondin12}, SN~1999aa \citep[from][]{blondin12}, and the normal \sneia\ SN~2005cf \citep{wang05cf} and SN~2003du \citep{stanishev03du}. We see that at maximum light SN~2012fr shows slightly weaker \SiII\ $\lambda$5972 and \SiII\ $\lambda$6355 than the normal \sneia, but higher velocity and stronger Ca H\&K absorption than SN~1999aa and the SN 1991T-like SN. At very early epochs ($-12$ days) SN~2012fr is very dissimilar to the extremely slow decliners, as it displays stronger absorption in \siline\ and the ``sulfur W,'' and it lacks the characteristic strong Fe absorption found in SN 1999aa/SN 1991T-like \sneia. The stronger absorption and higher velocities of the photospheric components of the \siline\ line and the lack of strong Fe absorption also argue against the possibility of SN~2012fr being a SN 1999aa/SN 1991T-like \snia\ with HVFs superposed on its spectrum. Thus, while SN~2012fr shares a number of characteristics with these very slowly declining \sneia, it does not exhibit sufficient spectroscopic similar to be classified as a member of this peculiar \snia\ subclass.

\section{Discussion}
\label{sec:discussion}
The spectra of SN~2012fr exhibit several noteworthy characteristics: (1) a very narrow velocity width in the photospheric absorption lines for phases later than one week after maximum light; (2) high-velocity \siline\ and \canir\ features which could be cleanly decoupled from the lower velocity photospheric component; (3) a clear plateau in the \siline\ and the \canir\ velocities, extending out until +39 days; (4) Si absorption-line strengths placing it on the borderline between ``shallow silicon'' and ``core normal'' classes; and (5) Si velocity placing it on the borderline between ``normal'' and ``high-velocity'' \sneia.

In this section we first reflect on what the \siline\ velocity behavior of SN~2012fr can reveal about interpretation of this velocity evolution in other \sneia\ (\S~\ref{sec:silicon_velocities}). We then consider how the aforementioned observational characteristics of SN~2012fr inform us about the probable nature of its explosion (\S~\ref{sec:explosion}). Finally, we consider the viability of SN~2012fr as a fundamental calibrator for measuring the Hubble constant (\S~\ref{sec:fund_calib}).

\subsection{SN~2012fr and Velocity Evolution of Other \sneia}
\label{sec:silicon_velocities}
One of the most noteworthy features of the spectra of SN~2012fr was the extremely clear distinction between the low-velocity photospheric \siline\ and an HVF. Clear identification of two absorption minima in the \siline\ line has only been previously observed convincingly in SN~2009ig \citep{marion13}, but HVFs may manifest themselves more subtly in the line profiles of other \sneia. This may then have an impact on the measurement of the \siline\ line velocity.

\begin{figure}
\begin{center}
\includegraphics[width=0.45\textwidth]{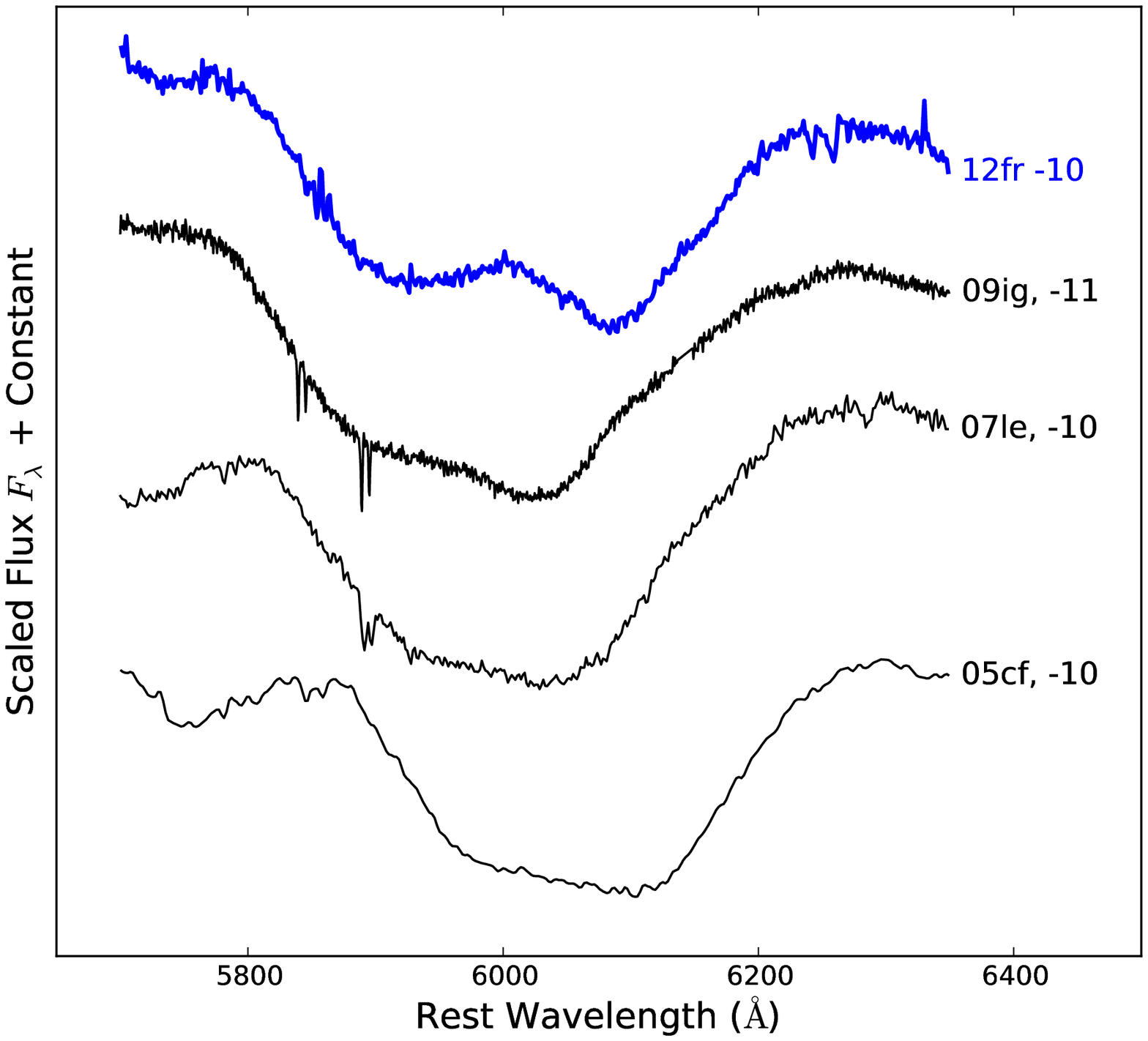}
\caption{\siline\ feature in the $-10$ day spectrum of SN~2012fr (blue) vs. early spectra of other \sneia\ with likely \siline\ HVFs: SN~2009ig \citep{foley09ig, marion13}, SN~2007le \citep[from][]{blondin12}, and SN~2005cf \citep{wang05cf}.}
\label{fig:early_snid}
\end{center}
\end{figure}

To inspect how common early HVF behavior is in \sneia, we searched for \sneia\ having spectra similar to the $-10$ day spectrum of SN~2012fr by employing the SN identification \citep[SNID;][]{snid} code. The top matches were, as expected, spectra of \sneia\ at about 10 days before maximum light, many of which exhibited a broad boxy absorption profile in the \siline\ line. In Figure~\ref{fig:early_snid} we plot the \siline\ profile of three notable matches -- SN~2009ig \citep{foley09ig, marion13}, SN~2007le \citep[from][]{blondin12}, and SN~2005cf \citep{wang05cf} -- all of which exhibit an absorption-line profile which seems difficult to explain with a single component, either Gaussian or P-Cygni. The \siline\ shape is best explained by a two-component profile like that observed in SN~2012fr, and this probability has been noted by previous authors \citep{mazzali01b, wang05cf, foley09ig}.

\begin{figure}
\begin{center}
\includegraphics[width=0.45\textwidth]{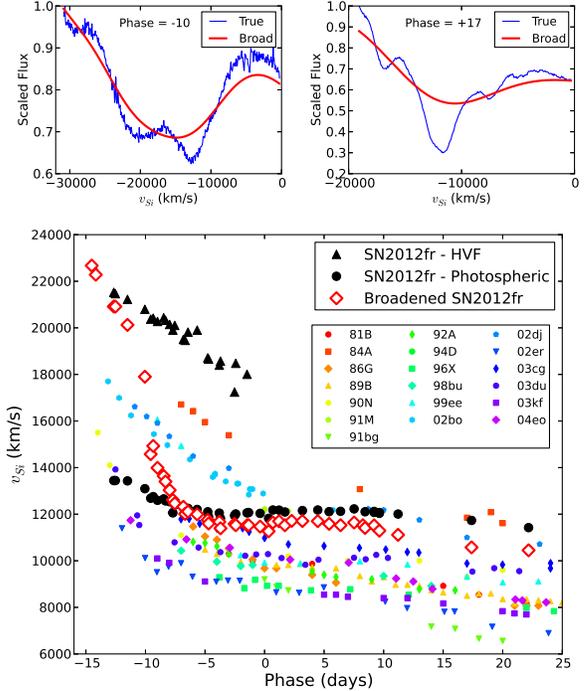}
\caption{Top panels: Observed spectra (blue) of SN~2012fr at $-10$ and +17 days, compared to spectra broadened by 3500 km$^{-1}$ (red). Bottom: Same as Figure~\ref{fig:all_vsi}, but with velocity measurements of the broadened SN~2012fr spectra shown as open red diamonds.}
\label{fig:broad_vsi}
\end{center}
\end{figure}

A possible consequence of early high-velocity features in \sneia\ could be an overestimate of the velocity gradient in the \siline\ line. To test this possibility, we convolved our spectra of SN~2012fr with a Gaussian filter of width $\sigma=$ 3500~km~s$^{-1}$ (FWHM $\approx$ 6700~km~s$^{-1}$) and measured the velocity minimum of the \siline\ line. Several examples of the broadened spectra, along with the velocity evolution measured from these spectra, are shown in Figure~\ref{fig:broad_vsi}. During the early epochs when the \siline\ HVF is clearly distinct in the observed spectra, our convolved spectra show a single broad absorption feature whose velocity declines smoothly from $\sim$22,000~km~s$^{-1}$ to $\sim$12,000~km~s$^{-1}$ over about 6 days (velocities were almost exactly the pEW-weighted mean of the values from Table~\ref{tab:vsi_fits}). 

Though this implied velocity gradient is much higher than that observed in any other \sneia, it illustrates the fact that multiple distinct components blended in velocity space could masquerade as a single rapidly evolving component. Inspection of the shape of the \siline\ absorption profile in HVG \sneia\ may provide insight into the possible impact of high-velocity features on the measured velocity gradients of these SNe.

Finally, we note that the long velocity plateau observed in SN~2012fr may not be unique, but instead may have been missed in other \sneia\ where the \siline\ line is blended with the neighboring Fe lines at late times. Inspection of the full line complex in spectra later than +20 days (see Figure~\ref{fig:time_series}) shows that the mean wavelength of the absorption lines near \siline\ becomes redder, due to multiple Fe lines appearing to the red of \siline, while the individual lines remain at constant velocity. This behavior may be hidden in other \sneia\ having broader lines, or it may be misinterpreted as a decrease in the line velocity. Indeed, our test of the velocity-broadened spectra of SN~2012fr revealed a slight decrease in the broadened velocity minimum starting at about +10 days. Future modeling of the SN~2012fr spectra released here may be able to address this question in more detail by identifying the lines neighboring  \siline\ and how their relative strengths evolve with time.

\subsection{The Nature of the SN~2012fr Explosion}
\label{sec:explosion}
The key observational features of SN~2012fr -- long Si and Ca velocity plateau, narrow absorption features, and early HVFs -- provide critical clues to the nature of its explosion. We will argue here that these observations indicate a thin region of partial burning products which produce the observed narrow line widths and IME velocity plateau, and which may be indicative of stratification in the ejecta. We then speculate on the origin of the HVFs, particularly in the context of recent proposals for surface He-shell burning.

\subsubsection{Ejecta Stratification}
The narrow velocity widths of the absorption profiles in SN~2012fr have enabled us to unambiguously track the velocity of the \siline\ line to very late times, revealing a velocity plateau at $\sim$12,000~km~s$^{-1}$ until at least the final epoch of observation at +39 days.
\sneia\ with such low velocity gradients have been observed before \citep[e.g., the ``LVG'' subclass from][]{benetti05}.
Notably, several \sneia\ have been found to exhibit a short-lived (i.e., few weeks duration) velocity plateau, such as the super-Chandrasekhar candidates presented by \citet{scalzo12}, as well as SN~2005hj \citep{quimby07} which also demonstrated narrow velocity width.

The velocity plateaus for these other \sneia\ were explained in in terms of explosion models featuring density enhancements at a particular velocity, which form at the reverse shock of an interaction between the SN ejecta and overlying material.  Examples include ``tamped detonations'' and ``pulsating delayed detonations,'' such as the DET2ENV and PDD series models of \citet{kmh93}. In a tamped detonation, the SN ejecta interact with a compact envelope of H-poor material, such as the C-O envelope which might remain after a double-degenerate merger \citep{fryer10,shen12}; the interaction freezes out quickly and the shock structure expands homologously thereafter. In a pulsating delayed detonation, the initial deflagration phase quenches and the WD progenitor undergoes a strong pulsation, causing the outer layers to contract.  The contraction then reignites C and eventually results in a transition to a detonation, in which a shock forms at the interface between the expanding inner layers and contracting outer layers.

Regardless of how it forms, the dense layer in these models remains optically thick for some time, resulting in a plateau in the \SiII\ velocity. Our analysis of \FeII\ lines in SN~2012fr, however, suggests that \SiII\ detaches from the photosphere between day +8 and day +11, disfavoring a pure density enhancement as the cause of the \SiII\ plateau in SN~2012fr. Additionally, in a tamped detonation one would expect that material above the density-enhanced layer would correspond primarily to the unburned envelope; thus the \SiII\ HVF we observe in SN~2012fr is difficult to explain in this scenario.

Rather than indicating a pure density enhancement, the velocity plateau in \siline\ may be evidence of \SiII\ being confined to a narrow region in velocity space. This possibility is reinforced by the analogous velocity plateau in the \canir\ (\S~\ref{sec:ca_nir_triplet}) as well as the slightly lower velocity of the \FeII\ lines at late times (\S~\ref{sec:fe_group}). This layering, along with the narrowness of the absorption features, indicates a probable stratification of the progress to complete nuclear burning in the ejecta. These properties are common in scenarios where the IMEs are produced mainly in a detonation phase \citep[see, e.g., the recent work by][]{seitenzahl13} as compared to deflagration scenarios, which result in significantly increased mixing in the ejecta.

Corroboration of stratification in the ejecta of SN~2012fr will be aided by nebular spectroscopy. Specifically, if the IMEs are distributed in a spherically symmetric shell, these should manifest themselves as double-peaked nebular lines such as the argon lines astutely observed by \citet{gerardy07} in the mid-IR nebular spectrum of SN~2005df. Conversely, if the IMEs are distributed asymmetrically, as discerned from spectropolarimetric observations by \citet{maund12fr}, this may be revealed in asymmetry of the nebular IME line profile (unless, of course, the asymmetric geometry is not oriented favorably). However, IME nebular lines are challenging to observe, and most lines in nebular spectra arise from material burned to full nuclear statistical equilibrium (Fe and Ni). As \citet{maund12fr} note, a global asymmetry of SN~2012fr would produce velocity shifts of the nebular Fe lines, following the prediction of \citet{maeda10}.

\subsubsection{High-Velocity Features}
The HVFs in both Si and Ca also shed light on the nature of the SN~2012fr explosion.
HVFs in the \canir\ appear to be a very common, perhaps even ubiquitous, feature in early-time \snia\ spectra \citep[e.g.,][]{mazzali05b}. HVFs in the \siline\ line have also been observed in many \sneia\ \citep[e.g.,][]{wang05cf, foley09ig, blondin12, bsnip2, marion13}. Positive identification of HVFs in other lines has been more challenging \citep[though see][for a thorough inspection of HVFs in multiple atomic species in SN~2009ig]{marion13}, so much effort has been focused on investigating the \canir\ and \siline\ HVFs.
Both the geometric distribution and physical origin of the material responsible for these HVFs are active areas of investigation. 

 \citet{tanaka06} posited that the HVFs could be due to patches of material outside the nominal photosphere, and the relative strength of the photospheric and HVFs is due to the relative ``covering fraction'' of the outer layer of material. If the HVFs originated from patchy layers of material, they would likely lack spherical symmetry. \citet{maund12fr} recently presented spectropolarimetry of SN~2012fr and argued that the relatively high degree of polarization in the HVFs is inconsistent with spherically symmetric geometry. Such asymmetry in HVFs has been inferred from spectropolarimetry of other \sneia\ as well \citep[for a review, see][]{ww08}.

The physical origin of HVF material remains unclear. It has been suggested that HVFs could be due to circum-stellar material \citep[CSM;][]{gerardy04, mazzali05a}, and for some \sneia\ CSM models have yielded favorable agreement with early-time \snia\ spectra with HVFs \citep{altavilla07, tanaka08}. In this scenario, absorption by \SiII\ and \CaII\ may be enhanced if the CSM is partially enriched in H \citep[CSM with $X(H) < 0.3$ would not produce a detectable \ha\ feature;][]{tanaka08}, thereby favoring a lower ionization state for these ions. However, significant absorption by Ca and Si across a range of velocities in SN~2012fr is indicative of these features being produced by material in (or on) the WD that undergoes partial nuclear burning during (or prior to) the SN explosion.

A possible explanation for the HVFs in SN~2012fr may be the detonation of He-rich material at the surface of the exploding WD. If a He layer of sufficiently low density is consumed by a detonation, the low density prevents burning to Fe-group elements but can be sufficient to produce Si and Ca (K. Nomoto 2012, private communication). In the single-degenerate Chandrasekhar-mass scenario, a He layer may be present following accretion from a binary companion. Explosive burning in a surface He layer is also a generic feature in the double-detonation scenario \citep{fink07, fink10, kromer10, sim10, sim12}, where the detonation of the He layer induces a detonation near the core of the WD progenitor, and even in merger scenarios in which the CO WDs retain a small He-rich atmosphere (R. Pakmor 2012, private communication). Recent work by \citet{townsley12} showed that He-shell burning could produce significant amounts of Ca, especially if the surface layer dredges up C from the WD surface through convection. However, one drawback of the He-shell scenario is that significant amounts of unburned He remain in most simulations, and while the remaining He should produce clear signatures in \snia\ spectra \citep{mazzalilucy98}, these features are not actually observed in \sneia.

Another possibility is that Si and Ca abundances are enhanced in the outer layers of a WD due to surface He burning that occurred prior to the SN event. This might arise during a surface He flash \citep[e.g.,][]{nomoto13} that manifests itself as a recurrent nova. This He-shell flash could produce relatively high Si and Ca abundances in the outer layers of the WD, or it might also ejecta partial burning products. These could either fall back onto the surface of the WD (again providing enrichment of the outer WD layers prior to explosion), or perhaps enrich part of the CSM into which the SN ejecta subsequently expand, and in principle the ejected material might be directly detected as CSM interaction of the SN \citep[e.g., as in PTF11kx;][]{dilday12}.

Thus, it is difficult at this time to constrain the exact origin of HVFs in \sneia, including SN~2012fr, but our data provide important observational constraints. HVFs in both \siline\ and the \canir\ were observed from two weeks before maximum light until they faded to obscurity at roughly $-2$ and +10 days, respectively. These HVFs were cleanly distinguished from the thin photospheric shell, and showed strong velocity gradients consistent with a receding photosphere. The velocity and absorption-strength evolution of these HVFs, along with the geometry implied from spectropolarimetry \citep{maund12fr}, provide important constraints for any model which posits an explanation for these features.

\subsection{SN~2012fr and the Hubble Constant}
\label{sec:fund_calib}
SN~2012fr occurred in NGC~1365, one of the galaxies included in the \emph{HST} Key Project on the extragalactic distance scale \citep{freedman01, silbermann99}. This fortuitous situation has made it a leading candidate to contribute toward measurement of the peak luminosity of \sneia. Such information is critical for determining the Hubble constant, and in the recent H$_0$ measurement by \citet{riess11} only eight \sneia\ with good light curves had independent distance measurements from observations of Cepheid variable stars in their host galaxies. Due to the existing Cepheid data for NGC~1365, SN~2012fr stands poised to be added to this sample. Because much of the effort in studying potential spectroscopic subclasses of \sneia\ has been aimed at improving the standardization of their luminosities, it is valuable to investigate where the fundamental calibrator \sneia\ sit in this context. Thus, in this section we inspect where the existing and probable future members of the Hubble-constant sample, including SN~2012fr, reside in the new \snia\ spectroscopic classification schemes.

\begin{table*}
\begin{center}
\caption{Spectroscopic Classes of $H_0$ Fundamental Calibrators}
\label{tab:fund_calib}
\begin{tabular}{llccccrccl}
\hline
SN & Host & $d^a$ & $\Delta m_{15}(B)$ & pEW(5972) & pEW(6355) & $v_{\rm Si}$ & Branch & Wang  & References \\
   &      & (Mpc) &  (mag)                 & (\AA)       & (\AA)       & (km\,s$^{-1}$)   & Class  & Class & \\
\hline
SN1981B  & NGC 4536 & 14.8 & $1.07 \pm 0.09$ &  20.0 & 129.0 & 13754 & BL & HV & 1,2,3 \\
SN1990N  & NGC 4639 & 21.6 & $1.00 \pm 0.03$ &  10.9 &  87.1 &  9352 & CN & N  & 4,5 \\
SN1994ae & NGC 3370 & 26.6 & $0.96 \pm 0.04$ &   7.4 &  81.6 & 10979 & CN & N  & 5 \\
SN1995al & NGC 3021 & 30.5 & $0.87 \pm 0.04$ &  15.0 & 112.9 & 12149 & BL & HV & 5 \\
SN1998aq & NGC 3982 & 22.5 & $1.11 \pm 0.04$ &  11.1 &  77.1 & 10796 & CN & N  & 5 \\
SN2002fk & NGC 1309 & 32.5 & $1.13 \pm 0.03$ &  10.3 &  75.7 & 10057 & CN & N  & 5 \\
SN2007af & NGC 5584 & 22.4 & $1.04 \pm 0.01$ &  17.0 & 105.2 & 10969 & BL & N  & 5 \\
SN2007sr & NGC 4038 & 21.7 & $1.13 \pm 0.06$ & \nodata & \nodata & \nodata & \nodata & \nodata & 6 \\
\hline
SN1998dh & NGC 7541         & 36.7 & $1.17 \pm 0.06$ &  10.1 & 121.8 & 12091 & BL & HV & 5 \\
SN2001el & NGC 1448         & 15.9 & $1.13 \pm 0.04$ &  12.0 &  93.0 & 11321 & CN & N  & 7,8 \\
SN2003du & UGC 9391         & 26.1 & $1.07 \pm 0.06$ &   1.9 &  88.8 & 10527 & CN & N  & 5,9 \\
SN2005cf & MCG -01-39-3$^b$ & 26.4 & $1.05 \pm 0.03$ &   6.5 &  99.1 & 10352 & CN & N  & 10,5 \\
SN2006D  & MCG -01-33-34    & 34.9 & $1.35 \pm 0.05$ &  21.4 &  94.4 & 10833 & CN & N  & 5 \\
SN2009ig & NGC 1015         & 35.9 & $0.89 \pm 0.02$ &   4.7 &  79.9 & 13400 & CN & HV & 11, 12, 5 \\
SN2011fe & M101             &  6.7 & $1.07 \pm 0.06$ &  15.1 & 101.4 & 10331 & CN & N  & 13, 14 \\
\hline
SN2012fr & NGC 1365 & 17.9 & $0.80 \pm 0.01$ &  3.9 &  66.5 & 12037 & SS/CN & HV/N & -- \\
\hline
\end{tabular}
\end{center}
$^a$ Distances from Cepheids for the \citet{riess11} sample,
\citet{kennicutt98m101} and \citet{freedman01} for M101, 
\citet{silbermann99} and \citet{freedman01} for NGC~1365, 
and redshift using H$_0=73.8$~km\,s$^{-1}$\,Mpc$^{-1}$ for others. \\
$^b$ Tidal bridge between MCG --01-39-3 and MCG --01-39-2 (NGC 5917); see \citet{wang05cf}. \\
References: (1) \citet{schaefer95}, (2) \citet{branch83}, (3) \citet{branch09}, 
(4) \citet{ganesh10}, (5) \citet{blondin12}, 
(6) \citet{hicken12}, (7) \citet{krisciunas03}, (8) \citet{wang03}, (9) \citet{stanishev03du},
(10) \citet{wang05cf}, (11) \citet{foley09ig}, (12) \citet{marion13}, 
(13) \citet{vinko12}, (14) \citet{parrent12}.
\end{table*}

In Table~\ref{tab:fund_calib} we summarize the light-curve decline rate and spectroscopic indicator data for the current sample of H$_0$ calibrators from \citet{riess11}, as well as a sample of seven additional \sneia\ which are strong candidates to be added to the H$_0$ sample, along with SN~2012fr. Selection criteria for the ``likely'' sample include (i) low reddening of the SN from its host, i.e., $A_V \lesssim 0.5$ mag, (ii) high sampling of the SN light curve including pre-maximum data, (iii) no spectroscopic peculiarity of the SN, (iv) distance modulus of the host of $\mu \lesssim 32.8$ mag, and (v) sufficiently low inclination to avoid crowding of Cepheids (L. Macri \& A. Riess 2012, private communication). The seven new \sneia\ which likely satisfy the above criteria are SN~1998dh, SN~2001el, SN~2003du, SN~2005cf, SN~2006D, SN~2009ig, and SN~2011fe.

We note here that for our selection criteria, spectroscopic peculiarity is explicitly defined as similarity to the peculiar \snia\ subclasses defined by their respective prototypes SN~1991bg, SN~1991T, SN~1999aa, and SN~2002cx. While SN~2012fr exhibits some rare behaviors, such as very narrow lines and a late velocity plateau in the IMEs, these characteristics are all seen in other \sneia\ (though not previously in this combination). Most importantly, the composition of SN~2012fr inferred from its spectral features is consistent with that of normal \sneia\ (see Figure~\ref{fig:99aa_comparison}), and the more subtle characteristics of its absorption features fall within the range of typical \snia\ behavior.

\begin{figure}
\begin{center}
\includegraphics[width=0.45\textwidth]{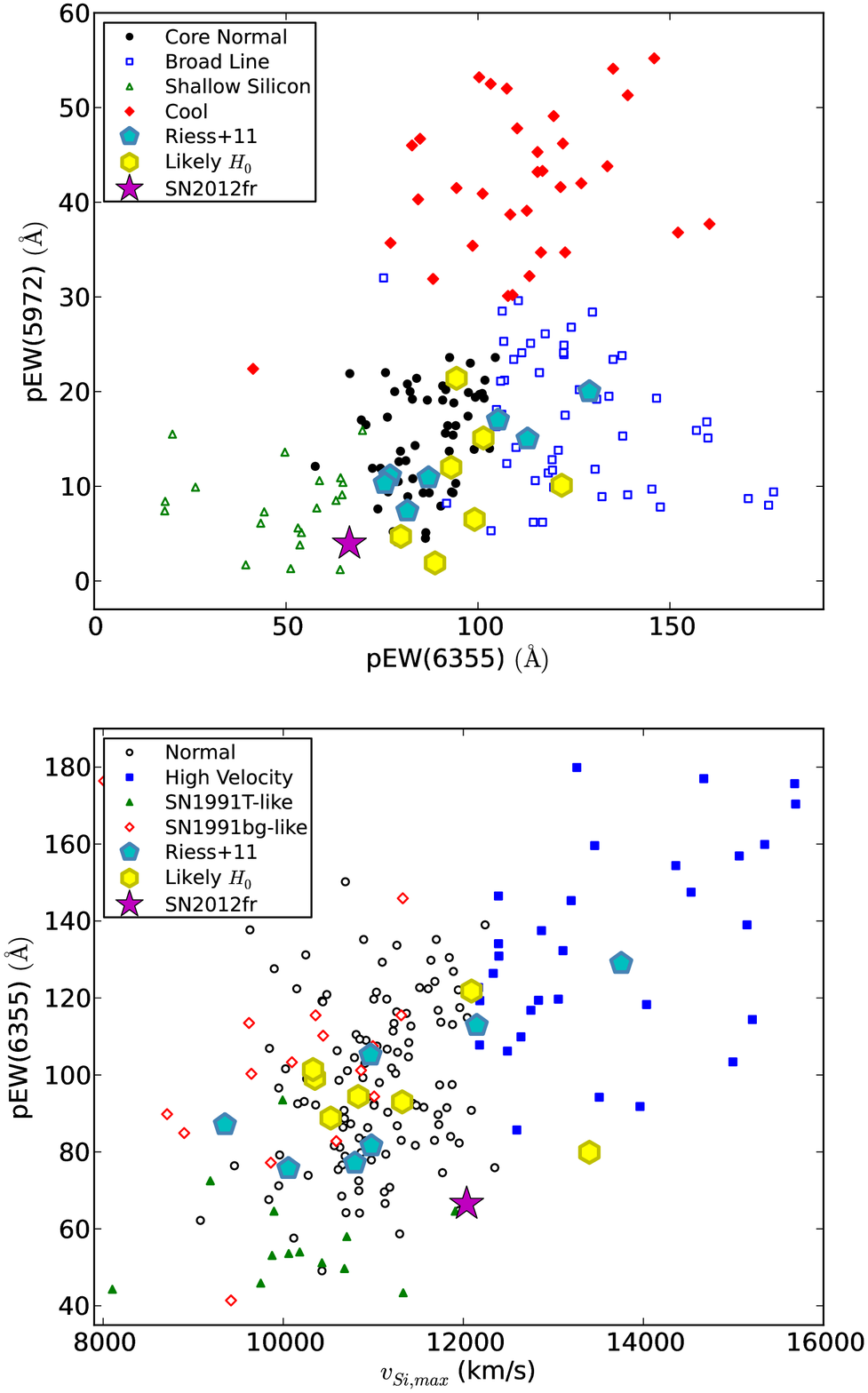}
\caption{Current members of the H$_0$ fundamental calibrator sample (cyan pentagons) and likely new additions to the sample (yellow hexagons), along with SN~2012fr (purple star), compared to other \sneia\ on the \citet{branch09} diagram (top) and the \citet{wang09} diagram (bottom). Other \snia\ subclasses are denoted as in Figure~\ref{fig:spec_inds}.}
\label{fig:fund_calib}
\end{center}
\end{figure}

In Figure~\ref{fig:fund_calib} we plot the location of existing and likely future H$_0$ calibrator \sneia\ along with SN~2012fr and the same comparison sample from Figure~\ref{fig:spec_inds}. The H$_0$ calibrator sample is not composed exclusively of \sneia\ falling in the ``core normal'' spectroscopic class of \citet{branch09} or the ``normal'' velocity class of \citet{wang09}. SN~2012fr would have the slowest light-curve decline rate and shallowest Si absorption of the H$_0$ sample, but not the highest velocity.
This prompts the question of what degree of ``normality'' is required for a \snia\ to be used for cosmology. The answer will likely require further study of \snia\ spectroscopic diversity and its impact on \snia\ luminosities.

\section{Conclusions}
\label{sec:conclusions}
We present optical spectra of SN~2012fr, a relatively normal Type Ia supernova which exhibits several interesting features. These include the distinct presence of high-velocity features in both the \siline\ line and \canir, as well as a long-lived velocity plateau at late epochs in both lines. These behaviors were made more clear by the extremely narrow velocity width of the photospheric absorption-line profiles. 
We show that SN~2012fr has \SiII\ velocities and absorption strengths which place it near the boundary between the ``shallow silicon'' and ``core normal'' spectroscopic classes defined by \citet{branch09}, and on the boundary between normal and ``high-velocity'' \sneia\ as defined by \citet{wang09}.

SN~2012fr exhibits a very slow decline rate ($\Delta m_{15}(B)=0.80\pm0.01$ mag; see Paper II), relatively shallow Si absorption at maximum light (pEW(6355) $= 66.5\pm15.5$~\AA), and a very low velocity gradient ($\dot v = 0.3 \pm 10.0$ km\,s$^{-1}$\,day$^{-1}$). All of these characteristics are common in the very luminous SN~1991T-like and SN~1999aa-like \snia\ subclasses, but SN~2012fr has higher ejecta velocities and much stronger Si and Ca absorption at early epochs than most \sneia\ in these classes (see Figure~\ref{fig:99aa_comparison}). Thus, SN~2012fr likely represents the most luminous end of the normal \snia\ spectrum, and it may be a transitional object between normal and very luminous events.

In the modern era only three well-observed normal \sneia\ (SN~1981B, SN~2001el, SN~2011fe) have occurred in galaxies that are both suitable for Cepheid distances and closer than the host galaxy of SN~2012fr, NGC~1365. Furthermore, this galaxy has been extensively studied and was part of the \emph{HST} Key Project on the extragalactic distance scale. This makes SN~2012fr a prime candidate for measuring the peak luminosity of \sneia\ and thereby constraining the Hubble constant, and we show that SN~2012fr should provide an excellent complement to the existing and likely future sample of H$_0$ \sneia.

Individual \sneia\ such as SN~2012fr with extensive observational datasets can prove invaluable for better understanding the nature of \snia\ explosions. Here we already identify a number of tantalizing observational characteristics of this spectral time series, and we expect that future detailed modeling could reveal additional subtle insights.

\vskip0.2in
{\it Facilities:} 
\facility{ANU:2.3m (WiFeS)},
\facility{NTT (EFOSC2)},
\facility{NTT (SOFI)},
\facility{SALT (RSS)},
\facility{SAAO:1.9m (Grating Spectrograph)},
\facility{Shane (Kast Double spectrograph)}, 
\facility{Keck:I (HIRES)}

\vskip11pt
\scriptsize
{\em Acknowledgments:}
We are very grateful to the staff of RSAA and Siding Spring Observatory for their rapid replacement of a broken cooling pump for the \wifes\ red channel on November 1, particularly Graeme Blackman, Gabe Bloxham, Donna Burton, Harvey Butcher, Mike Ellis, Mike Fowler, Mike Petkovic, Annino Vaccarella, and Peter Verwayen.  We thank Bruce Bassett for arranging observations with the SAAO 1.9 m telescope, as well as Carl Melis, Michael Jura, Siyi Xu, Beth Klein, David Osip, Ben Zuckerman, and Barry Madore for contributing data. We are grateful to Stephane Blondin for providing his velocity-gradient values from the CfA SN~Ia sample, and to Ken Nomoto, Rudiger Pakmor, Lucas Macri, and Adam Riess for helpful discussions. We also thank Howie Marion for providing an advance copy of his paper on SN~2009ig, and for very helpful discussions. 

This research was conducted by the Australian Research Council Centre of Excellence for All-sky Astrophysics (CAASTRO), through project number CE110001020.  
Chris Lidman is the recipient of an Australian Research Council Future Fellowship (program number FT0992259).  
J.A. acknowledges support by CONICYT through FONDECYT grant 3110142, and by the Millennium Center for Supernova Science (P10-064-F), with input from 'Fondo de Innovaci\'on para la Competitividad, del Ministerio de Econom\'ia, Fomento y Turismo de Chile'. 
A.G.-Y. is supported by the EU/FP7 via an ERC grant.
F.B. acknowledges support from FONDECYT through Postdoctoral grant 3120227. F.B. and G.P. thank the Millennium Center for Supernova Science for grant P10-064-F (funded by ``Programa Bicentenario de Ciencia y Tecnolog\'ia de CONICYT'' and ``Programa Iniciativa Cient\'ifica Milenio de MIDEPLAN'').
S.B. is partially supported by the PRIN-INAF 2011 with the project ``Transient Universe: from ESO Large to PESSTO''.
Support for this research at Rutgers University was provided in part by NSF CAREER award AST-0847157 to S.W.J.
M.D.S. and F.T. acknowledge the generous support provided by the Danish Agency for Science and Technology and Innovation through a Sapere Aude Level 2 grant. 
E.Y.H. is supported by the NSF under grant AST-1008343.
A.V.F.'s group at U.C. Berkeley is supported by Gary and Cynthia Bengier, the Richard and Rhoda Goldman Fund, the Christopher R. Redlich Fund, the TABASGO Foundation, and NSF grant AST-1211916.

This work is based in part on observations collected at the European Organisation for Astronomical Research in the Southern Hemisphere, Chile, as part of PESSTO (the Public ESO Spectroscopic Survey for Transient Objects) ESO programs 188.D-3003 and 089.D-0305.
This paper also uses data obtained at the South African Astronomical Observatory (SAAO). Some observations were taken with the Southern African Large Telescope (SALT) as part of proposal ID 2012-1-RU-005 (PI: Jha).
Some of the data presented herein were obtained at the W. M. Keck Observatory, which is operated as a scientific partnership among the California Institute of Technology, the University of California, and NASA; the observatory was made possible by the generous financial support of the W. M. Keck Foundation.
Based in part on observations made with the Nordic Optical Telescope, operated on the island of La Palma jointly by Denmark, Finland, Iceland, Norway, and Sweden, in the Spanish Observatorio del Roque de los Muchachos of the Instituto de Astrofisica de Canarias. 
For their excellent assistance, we are grateful to the staffs of the 
many observatories where we collected data.

This research has made prodigious use of the NASA/IPAC Extragalactic Database (NED), which is operated by the Jet Propulsion Laboratory, California Institute of Technology, under contract with NASA.
It has also made use of NASA's Astrophysics Data System (ADS), the CfA 
Supernova Archive (funded in part by NSF grant AST-0907903), and the 
Central Bureau for Astronomical Telegrams (CBAT) list of SNe 
(http://www.cbat.eps.harvard.edu/lists/Supernovae.html).



\bibliographystyle{apj}
\bibliography{sn2012fr_spec}

\end{document}